\documentclass[%
 reprint,
 groupedaddress,
superscriptaddress,
 amsmath,amssymb,
 aps,
]{revtex4-1}

\usepackage{mathrsfs}
\usepackage{textcomp}
\usepackage{units}
\usepackage{array,mathtools}
\usepackage{stmaryrd}
\usepackage{scrextend}
\usepackage{nicefrac}
\usepackage{multirow}
\usepackage{makecell}
\usepackage{bm}
\newcolumntype{C}{>{$}c<{$}}
\AtBeginDocument{
\heavyrulewidth=.08em
\lightrulewidth=.05em
\cmidrulewidth=.03em
\belowrulesep=.65ex
\belowbottomsep=0pt
\aboverulesep=.4ex
\cmidrulesep=\doublerulesep
\cmidrulekern=.5em
\defaultaddspace=.5em
}
\usepackage{booktabs}
\usepackage[caption=false]{subfig}
\usepackage{graphicx}
\usepackage{dcolumn}
\usepackage{bm}
\usepackage{xspace}

\usepackage{xcolor}
\usepackage[normalem]{ulem}

\newcommand{\editor}[2]{%
  \expandafter\newcommand\csname #1note\endcsname[1]{%
    \textcolor{#2}{(\textbf{#1:} ##1)}}%
  \expandafter\newcommand\csname #1\endcsname[1]{%
    \textcolor{#2}{##1}}%
  \expandafter\newcommand\csname #1cancel\endcsname[1]{%
    \textcolor{#2}{\sout{##1}}}%
  \expandafter\newcommand\csname #1change\endcsname[2]{%
    \textcolor{#2}{\sout{##1} ##2}}%
  \newenvironment{#1text}{\color{#2}}{\color{black}}
}

\begin{document}

\title{Noncollinear DFT+\textit{U} and Hubbard parameters with fully-relativistic ultrasoft pseudopotentials}

\author{Luca Binci}
\affiliation{Theory and Simulation of Materials (THEOS), and National Centre for Computational Design and Discovery of Novel Materials (MARVEL), École Polytechnique Fédérale de Lausanne, CH-1015 Lausanne, Switzerland}
\author{Nicola Marzari}
\affiliation{Theory and Simulation of Materials (THEOS), and National Centre for Computational Design and Discovery of Novel Materials (MARVEL), École Polytechnique Fédérale de Lausanne, CH-1015 Lausanne, Switzerland}
\affiliation{Laboratory for Materials Simulations, Paul Scherrer Institut, 5232 Villigen PSI, Switzerland}
\date{\today}

\begin{abstract}
The magnetic, noncollinear parametrization of Dudarev's DFT+$U$ method is generalized to fully-relativistic ultrasoft pseudopotentials. We present the definition of the DFT+$U$ total energy functional, and the calculation of forces and stresses in the case of orthogonalized atomic orbitals defining the localised Hubbard manifold, where additional contributions arising from the derivative of the inverse square root of the overlap matrix appear. We further extend the perturbative calculation of the Hubbard $U$ parameters within density-functional perturbation theory to the noncollinear relativistic case, by exploiting an existing and recently developed theoretical approach that takes advantage of the time-reversal operator to solve a second Sternheimer equation. We validate and apply the new scheme by studying the electronic structure and the thermodynamics of the binary compounds EuX (where X = O, S, Se, Te is a chalcogen atom), as representative simple crystals, and of the pyrochlore Cd$_2$Os$_2$O$_7$, representative of a more structurally complex oxide.   
\end{abstract}

\maketitle
\section{introduction}
Simulating electronic properties of materials containing transition-metal or rare-earth atoms has always been a considerable challenge for first-principles methods based on density-functional theory (DFT). The reasons are twofold: strongly localized electrons can display multi-determinant character/strong correlations, that are not captured by approximate DFT, and they give rise to large self-interaction errors \cite{perdew_self-interaction_1981}. One of the most widely adopted scheme to improve the electronic description of systems hosting localized electrons is DFT+\textit{U} \cite{anisimov_band_1991,anisimov_first-principles_1997,himmetoglu_hubbard-corrected_2014}. The approach consists in augmenting the DFT total energy functional with a corrective term aiming to improve the description of \textit{d}- or \textit{f}-electrons: $E_{\mathrm{DFT}+U}=E_\mathrm{DFT}+E_U$. Rather than addressing electronic correlations, DFT+$U$ corrects the large self-interaction errors that are prevalent in systems that also display strong correlations \cite{kulik_density_2006}. In fact, the main effect of the $E_U$ correction enforces an integer-like electronic occupation in the target localized manifold, or in more modern terminology, a piece-wise linearity of the total energy with respect to the occupation of such a manifold \cite{anisimov_density-functional_1993,cococcioni_linear_2005}. It has been shown to be quite successful to describe the magnetic and insulating properties of oxides \cite{cococcioni_linear_2005,agapito_reformulation_2015} and molecules \cite{kulik_density_2006,kulik_self-consistent_2008} containing $3d$ transition-metals. Elements belonging to the same block of the periodic table but in the 5th and 6th rows have frontier $4d$ and $5d$ orbitals, which are spatially more extended and therefore less localised and magnetic than the $3d$ ones; for this reason the +$U$ correction is less frequently adopted to these ions. On the other hand, increasing the atomic number also increases the strenght of the relativistic effects, i.e. the spin-orbit coupling (SOC), which scales as $Z^4$ for hydrogenic atoms ($Z$ being the atomic number), and as $Z^2$ (due to the shielding effect) in the external shells of many-electron atoms \cite{pyykko_relativistic_1988}. As a consequence, in less explored but very interesting cases, it is the combined effect of electronic localization and SOC that determines the nature of the ground state. Relevant examples are the strontium iridate Sr$_2$IrO$_4$ \cite{kim_novel_2008}, pyrochlore iridates \cite{wan_topological_2011}, americium monochalcogenides  \cite{zhang_actinide_2012}, and the osmium-based double perovskite Ba$_2$NaOsO$_6$~\cite{erickson_ferromagnetism_2007}.

In this paper we introduce a generalization of the DFT+$U$ method to deal with noncollinear magnetic structures and SOC -- already described in \cite{dudarev_parametrization_2019,tancogne-dejean_self-consistent_2017} -- and extend it to a fully-relativistic (FR) ultrasoft pseudopotentials (US-PP) formalism \cite{vanderbilt_soft_1990,corso_spin-orbit_2005}. In doing this, we exploit Smogunov's implementation of Liechtenstein's DFT+$U$ functional with FR US-PP \cite{giannozzi_advanced_2017,liechtenstein_density-functional_1995} (which is currently restricted to total energy calculations only, without structural optimization or determination of Hubbard parameters), and then adapt it to Dudarev's parametrization of DFT+$U$ (which consists in the neglect of high-order multipolar terms in the Coulomb interaction tensor within Liechtenstein's DFT+$U$ \cite{cococcioni_linear_2005}). In the same PP framework, we extend the former implementation by deriving the expressions of forces and stresses, also when orthogonalized atomic orbitals are used as Hubbard projectors, thus requiring the inclusion of contributions coming from the derivatives of the inverse square root of the overlap matrix between different site-centered atomic-like orbitals. Finally, exploiting density-functional perturbation theory (DFPT) \cite{baroni_phonons_2001}, we present a scheme to calculate the Hubbard $U$ parameters from first-principles in the case of broken time-reversal symmetry (induced by the noncollinear magnetism) and including SOC in full at the self-consistent level. In doing so, we leverage the recent approach introduced in Ref. \cite{urru_density_2019} for the calculation of phonon frequencies in the presence of SOC and noncollinear magnetism -- that exploits the solution of two independent Sternheimer equations: a standard one and a time-reversed one, in order to obtain the first-order wavefunctions entering in the linear-response problem. In this way, the entire formulation is internally consistent and completely free from any adjustable parameters. 

We then apply the theory we developed to the study of the series of binary europium monochalcogenides EuX (X = O, S, Se, Te) and of the more complex oxide osmium pyroclore Cd$_2$Os$_2$O$_7$. For the EuX compounds, we determine both electronic structure and thermodynamic properties, such as lattice parameter and bulk modulus, and quantify the effect of different parametrizations of the exchange-correlation functional on the final results. We also study the pressure dependence of the enthalpy, and evaluate the critical pressure above which a structural phase transition occurs, where the crystal changes from the $Fm\bar{3}m$ to the $Pm\bar{3}m$ space group. For Cd$_2$Os$_2$O$_7$, we study the electronic structure and energetics of different noncollinear magnetic orderings, and analyze how the magnetic patterns imposed, change the crystal structure following variable-cell structural relaxations. 

The paper is organized as follows: in Sec. \ref{theory_a} we introduce the DFT+$U$ functional in the FR US-PP scheme, and generalize the calculation of the forces and stress tensor to this framework, extending also to the case of orthogonalized atomic wavefunctions defining the localized Hubbard manifold. In Sec. \ref{theory_b} we discuss the theory concerning the perturbative calculation of the Hubbard $U$  parameters, generalized to the noncollinear case in absence of time-reversal symmetry. Finally, in Sec. \ref{applications} we validate the new approach by applying it to the study of EuX oxide and chalcogenides and Cd$_2$Os$_2$O$_7$.

\section{theoretical formulation\label{theory}}
\subsection{Total energy functional, forces and stresses\label{theory_a}}
Within DFT+$U$, the total energy DFT functional is modified by adding the so-called Hubbard corrective term. Dudarev's formulation of DFT+\textit{U} \cite{dudarev_electron-energy-loss_1998} provides a prescription for the Hubbard augmentation part that is rotationally invariant with respect to the localized manifold of interest. Here, we adopt the parametrization of the noncollinear Dudarev's functional as employed in \cite{dudarev_parametrization_2019,tancogne-dejean_self-consistent_2017}, and generalize it to the FR US-PP's scheme. In principle, when atomic states from a FR calculation are used to build the Hubbard occupation matrix \cite{cococcioni_linear_2005}, this latter should depend on the quantization-axis projection of the total angular momentum $\bm{J}=\bm{L}+\bm{S}$. However, to avoid to work in the basis of $\bm{J}$ and recast the formulation in a similar way as is done in electronic-structure codes for US-PP's \cite{vanderbilt_soft_1990}, we use the $j$-averaged radial parts of the atomic wavefunctions. The assumption underlying this procedure is that magnetism (mostly driven by the electronic exchange) is dominant compared to the splitting induced by SOC. In this way, the total energy functional acquires the spin-resolved form $E=E_\mathrm{DFT}+E_U$, where
\begin{equation}
     E_U=\sum_I\frac{U^I}{2}\sum_{m,\sigma}\bigg(n^{I\sigma\sigma}_{mm}-\sum_{m'\sigma'}n^{I\sigma\sigma'}_{mm'}\,n^{I\sigma'\sigma}_{m'm}\bigg)
\end{equation}
where $I$ labels the atomic site within the unit cell, and $m$ and $\sigma$ are the atomic magnetic quantum number and spin index, respectively. The occupation matrix $n_{mm'}^{I\sigma\sigma'}$ in the FR US-PP scheme reads \cite{timrov_self-consistent_2021}
\begin{gather}
n_{mm'}^{I\sigma\sigma'}=\sum_{i,\sigma_1,\sigma_2}\tilde{\theta}_i\,\big\langle\psi_{i}^{\sigma_2}\big|S^{\sigma_2\sigma'}\big|{\phi}^{I\sigma'}_{m'}\big\rangle\big\langle{\phi}^{I\sigma}_m\big|S^{\sigma\sigma_1}\big|\psi_{i}^{\sigma_1}\big\rangle\label{matr}\\
n^{I\sigma\sigma'}=\sum_m n_{mm}^{I\sigma\sigma'}=\frac{1}{2}\big(n^I\delta^{\sigma\sigma'}+\bm{\sigma}^{\sigma\sigma'}\cdot\bm{m}^I\big)
\end{gather}
where $\psi$ and $\phi$ are, respectively, the (pseudo) Kohn-Sham (KS) and localised Hubbard states, $n^I=\sum_{\sigma}n^{I\sigma\sigma}$ is the occupation and $\bm{m}^I=\sum_{\sigma\sigma'}n^{I\sigma\sigma'}\bm{\sigma}^{\sigma\sigma'}$ is the magnetization of the Hubbard manifold, and the index $i$ collectively represents the band and the quasimomentum indices. The 2$\times$2 ultrasoft (nonlocal) $S$ matrix $\langle\bm{r}_1| S^{\sigma\sigma'}|\bm{r}_2\rangle=S^{\sigma\sigma'}(\bm{r}_1,\bm{r}_2)$ is given by  
\begin{equation}
\begin{split}
S^{\sigma\sigma'}(\bm{r}_1,\bm{r}_2)=\;\,&\delta(\bm{r}_1-\bm{r}_2)\,\delta^{\sigma\sigma'}\\
&+\sum_{I,\nu,\nu'}Q^{I\sigma\sigma'}_{\nu\nu'}\,\beta^I_\nu(\bm{r}_1)\,\beta^{I}_{\nu'}(\bm{r}_2)^\ast,
\end{split}
\end{equation}
where $\nu$ is an index referring to the atomic quantum numbers of the $\beta$ projector functions pertaining to the US-PP, and 
\begin{equation}    Q^{I\sigma\sigma'}_{\nu\nu'}=\sum_{\nu_1,\nu_2,\sigma_1}f^{\sigma\sigma_1}_{\nu\nu_1}\,Q^I_{\nu_1\nu_2}\,f_{\nu_2\nu'}^{\sigma_1\sigma'}
\end{equation} 
are the FR augmentation charges. These lead to having orthonormal constraints $\sum_{\sigma\sigma'}\langle\psi^\sigma_i|S^{\sigma\sigma'}|\psi^{\sigma'}_j\rangle=\delta_{ij}$; see Ref. \cite{corso_spin-orbit_2005} for an in-depth discussion. The KS pseudo-wavefunctions $\psi_{i}^{\sigma}(\bm{r})=\langle \bm{r},\sigma|\Psi_i\rangle$ are defined as the $\sigma$-spin component of the KS spinor $|\Psi_i\rangle=\sum_\sigma|\psi_{i\sigma},\sigma\rangle$. The atomic pseudo-states $\phi^{I\sigma}_m=\phi^{I}_m$ are actually spin-independent (because of the $j$-averaging procedure), but for practical purposes we still define the spinor $|\Phi^I_m\rangle=\sum_\sigma |\phi_m^{I},\sigma\rangle$, for which $\phi_m^{I\sigma}(\bm{r})=\langle\bm{r},\sigma|\Phi^{I}_m\rangle$ (it will have nonzero spin-components as in the collinear case, and equal spatial parts). If these states are orthogonalized according to L{\"o}wdin's decomposition, an overlap matrix $O^{I\sigma I'\sigma'}_{mm'}=\langle\phi^{I\sigma}_{m}|S^{\sigma\sigma'}|\phi^{I'\sigma'}_{m'}\rangle$ needs to be introduced and the Hubbard atomic-like states become
\begin{equation}
 \phi^{I\sigma}_{m}(\bm{r})\rightarrow \varphi^{I\sigma}_{m}(\bm{r})=\sum_{I',m',\sigma'}\big(O^{-\nicefrac{1}{2}}\big)_{m'm}^{I'\sigma'I\sigma}\,\phi^{I'\sigma'}_{m'}(\bm{r}).\label{ortho-atomic}
\end{equation}
The former set of equations completely defines our total energy functional.

For the calculation of forces and stresses, the Hellmann-Feynamn theorem applies (provided the Hubbard $U$ is considered independent of displacements and strain); as a consequence, the variation of $E_U$ with respect to a parameter $\mu$ ($\mu=\bm{u}_I$, the atomic displacement, for forces, and $\mu=\bm{\varepsilon}$, the strain, for the stress tensor) involves only the variations of the ultrasoft matrix and the $\phi^{I\sigma}_m(\bm{r})$ functions:
\begin{equation*}
    \frac{\partial E_U}{\partial\mu}=\sum_I\frac{U^I}{2}\sum_{\{m\},\{\sigma\}}\Big(\delta_{mm'}^{\sigma\sigma'}-2n^{I\sigma\sigma'}_{mm'}\Big)\,\frac{\partial n^{I\sigma'\sigma}_{m'm}}{\partial\mu}
\end{equation*}
in which the terms to be differentiated, according to Eq.~(\ref{matr}), are
\begin{gather*}
    \frac{\partial }{\partial\mu}\langle\psi_{i}^\sigma|S^{\sigma\sigma'}|\phi^{I\sigma'}_{m'}\rangle\rightarrow \langle\psi_{i}^\sigma|\bigg[\frac{\partial S^{\sigma\sigma'}}{\partial\mu}|\phi^{I\sigma'}_{m}\rangle+S^{\sigma\sigma'}\bigg|\frac{\partial \phi^{I\sigma'}_{m}}{\partial\mu}\bigg\rangle\bigg]\\
    \frac{\partial S^{\sigma\sigma'}}{\partial\mu}=\sum_{I,\nu\nu'}Q^{I\sigma\sigma'}_{\nu\nu'}\bigg(\,\bigg|\frac{\partial \beta^I_\nu}{\partial\mu}\bigg\rangle\big\langle\beta^I_{\nu'}\big|+\big|\beta^I_\nu\big\rangle\bigg\langle\frac{\partial\beta^I_{\nu'}}{\partial\mu}\bigg|\,\bigg);
\end{gather*}
in the case of orthogonalized atomic orbitals, the derivative of Eq. (\ref{ortho-atomic}) involves also the variation of the overlap matrix 
\begin{equation*}
\begin{split}
\frac{\partial\varphi^{I\sigma}_{m}(\bm{r})}{\partial\mu}=&\sum_{I',m',\sigma'}\bigg[\bigg(\frac{\partial O^{-\nicefrac{1}{2}} }{\partial\mu}\bigg)_{m' m}^{I'\sigma' I\sigma}\,\phi^{I'\sigma'}_{m'}(\bm{r})\\
    &+\big( O^{-\nicefrac{1}{2}}\big)_{m'm}^{I'\sigma' I\sigma}\,\frac{\partial\phi^{I'\sigma'}_{m'}(\bm{r})}{\partial\mu}\bigg].\label{orth}
\end{split}
\end{equation*}
The case of nonorthogonal atomic states is recovered by setting $O^{I\sigma I'\sigma'}_{mm'}=\delta^{II'}\delta^{\sigma\sigma'}\delta_{mm'}$; notably, in this latter case, when evaluating forces, the second term of the former equation reduces to $|\partial_\mu\phi^{I'\sigma'}_{m'}\rangle=|\partial_{\bm{u}_J}\phi^{I'\sigma'}_{m'}\rangle=\delta^{I'J}|\partial_{\bm{u}_J}\phi^{J\sigma'}_{m'}\rangle$, and still contains the sum of inter-orbital terms within the same atom due to saturation of the $m'$ and $\sigma'$ indexes with the overlap matrix. For the detailed expressions concerning how to evaluate in practice the derivatives of the atomic orbitals in reciprocal space, we refer to Ref. \cite{cococcioni_accurate_2010}, and for the variation of the $\beta^I_\nu$ projectors we refer to Ref. \cite{dal_corso_density-functional_2001}. In particular, the calculation of the derivative of the inverse square root of $ O^{I\sigma I'\sigma'}_{mm'}$ is carried out using the formal solution of the Lyapunov equation
\begin{equation}
\begin{split}
        \frac{\partial O^{-\nicefrac{1}{2}}}{\partial\mu} =-&\int_0^\infty dt\,\exp\big(-tO^{-\nicefrac{1}{2}}\big)\\
        &\times O^{-1}\bigg(\frac{\partial O}{\partial\mu}\bigg)\,O^{-1}\,\exp\big(-tO^{-\nicefrac{1}{2}}\big)\label{doverlap}
\end{split}
\end{equation}
where the integral can be performed analytically element by element of $\partial_\mu O^{-\nicefrac{1}{2}}$ \cite{timrov_pulay_2020}. The usefulness of Eq. (\ref{doverlap}) -- although computationally expensive for systems with many atoms (large supercells) -- is that it is exact and shows that for the calculation of $\partial_\mu O^{-\nicefrac{1}{2}}$ only the eigenvalues of $O$ and of its derivative are needed.

\subsection{First-principles determination of Hubbard parameters using perturbation theory\label{theory_b}}
Following the linear-response approach for the calculation of the Hubbard parameters introduced by Cococcioni and de Gironcoli \cite{cococcioni_linear_2005}, the Hubbard $U$ can be defined as 
\begin{equation}
U^I=\big(\chi_0^{-1}-\chi^{-1}\big)_{II},\label{hubbard::U}
\end{equation}
where $\chi \;(\chi_0)$ are the interacting (noninteracting) response matrices to a localised perturbation indirectly changing -- through a Legendre transform -- the localised orbital occupation. In the original scheme of Ref. \cite{cococcioni_linear_2005}, in order to have a localized perturbing potential, a finite-difference approach based on supercells was used. More recently, Timrov et al. \cite{timrov_hubbard_2018,timrov_self-consistent_2021} developed a more efficient scheme based on DFPT, according to which the supercell periodicity is substituted by a decomposition of the potential into monochromatic perturbations, with wave-vectors belonging to a $\bm{q}$-point grid of size commensurate to an equivalent supercell. Although in Refs. \cite{timrov_hubbard_2018,timrov_self-consistent_2021} time-reversal symmetry was not assumed -- because of the presence of a spin-polarized exchange-correlation potential -- the collinearity of the problem still allowed to recast the perturbative expansion as in the time-reversal-invariant case. However, it has been shown \cite{cao_ab_2018,gorni_spin_2018,urru_density_2019} that when noncollinear magnetism is present, it is necessary to solve at least two separate Sternheimer linear systems. This approach was originally developed for the calculation of spin-fluctuation spectra in a norm-conserving pseudopotential (NC-PP) framework: in Ref. \cite{cao_ab_2018} two Sternheimer equations were solved, one at $+\bm{q}$ and the other at $-\bm{q}$, while in \cite{gorni_spin_2018} the time-reversal operator $\mathcal{T}$ was used in the second Sternheimer equation. In Ref. \cite{urru_density_2019} the scheme employing $\mathcal{T}$ was further generalized to FR US-PPs, and used for the calculation of phonon dispersions in elemental magnetic metals.

The original extension of DFPT to US-PPs was developed by Dal Corso for the calculation of vibrational spectra \cite{dal_corso_density-functional_2001}, who showed how the change of orthonormalization constraint -- provided by the presence of the $S$ matrix -- changes the induced charge density; then, by the same author, it was generalised to the FR nonmagnetic case \cite{dal_corso_density_2007}. In this work we follow the strategy of Ref. \cite{urru_density_2019}, where the former DFPT was extended to the magnetic time-reversal symmetry broken case, and apply it to an external perturbation changing the orbital occupation. Indeed, to evaluate \emph{ab initio} the Hubbard $U$ parameter in the linear response method of Ref. \cite{cococcioni_linear_2005}, one uses the constrained DFT technique \cite{dederichs_ground_1984,hybertsen_calculation_1989}: from the total energy $E$, a new energy functional is introduced $\tilde{E}\{q\}=\min_{\rho,q}\big\{E[\rho]+\sum_I\alpha_I(n^I-q^I)\big\}$, where the orbital occupations $n^I$ are controlled by the variables $q^I$. The additional term $\sum_I\alpha_I(n^I-q^I)$ acts precisely as a perturbation changing the orbital occupation of the localised manifold, thanks to which variations of $\tilde{E}\{q\}$ with respect to $q^I$ can be acquired. The Hubbard $U$ parameter is then defined as the curvature of $\tilde{E}\{q\}$ with respect to the variables $q^I$. In practice a Legendre transform is performed, which shifts the $q^I$-dependence to the $\alpha^I$-dependence: $\mathcal{E}=\tilde{E}\{q\}+\sum_I\alpha_Iq^I$. The curvature of $\mathcal{E}$ with respect to $\alpha_I$ yields the $\chi$ and $\chi_0$ matrices, which are linked to the Hubbard $U$ by Eq. (\ref{hubbard::U}) \cite{cococcioni_linear_2005,timrov_hubbard_2018}. 

Introducing the projector $P^{I,\sigma\sigma'}_{\sigma_1\sigma_2}=\sum_{m}S^{\sigma_1\sigma}|\phi^{I\sigma}_m\rangle\langle\phi^{I\sigma'}_m|S^{\sigma'\sigma_2}$ -- in terms of which the external perturbation becomes $\sum_I\alpha_I\sum_{i,\{\sigma\}}\tilde{\theta}_i\langle\psi_{i}^{\sigma_1}|P^{I,\sigma\sigma}_{\sigma_1\sigma_2}|\psi_{i}^{\sigma_2}\rangle$ (the notation $\sum_{\{\sigma\}}$ stands for a sum over all the $\sigma$ indices appearing in the following expression) -- the first derivative of the total energy $\mathcal{E}=E+\sum_I\alpha_I\sum_{m\sigma}n^{I\sigma\sigma}_{mm}$ with respect to the external parameter $\alpha_I$, because of Hellmann-Feynman theorem, is $\partial_{I}\mathcal{E}=n^I=\sum_{i,\{\sigma\}}\tilde{\theta}_i\,\langle\psi_{i}^{\sigma_1}|P^{I,\sigma\sigma}_{\sigma_1\sigma_2}|\psi_{i}^{\sigma_2}\rangle$ (we use the notation $\partial_I\equiv\partial/\partial\alpha_I$). The second derivative gives the response matrix $\partial_I\partial_{I'}\mathcal{E}=\chi_{II'}$, where
\begin{equation}
\begin{split}
\chi_{II'}&=\sum_{m,\sigma}\frac{\partial n_{mm}^{I\sigma\sigma}}{\partial \alpha_{I'}}=\sum_{i,\{\sigma\}}\bigg\{\bigg[\frac{\partial \tilde{\theta}_i}{\partial\alpha_{I'}}\bigg]\big\langle\psi_{i}^{\sigma_1}\big|P^{I,\sigma\sigma}_{\sigma_1\sigma_2}\big|\psi_{i}^{\sigma_2}\big\rangle\\
&+\bigg(\tilde{\theta}_i\,\big\langle\psi_{i}^{\sigma_1}\big|P^{I,\sigma\sigma}_{\sigma_1\sigma_2}\bigg|\frac{\partial \psi_{i}^{\sigma_2}}{\partial\alpha_{I'}}\bigg\rangle+\mathrm{c.c.}\bigg)\bigg\}. 
\end{split}
\end{equation}
The first term accounts for the change of electronic occupation and for the intraband term, and it is present in metals for perturbations with the same periodicity of the underlying lattice \cite{baroni_phonons_2001}. This latter presents no relevant difference compared to the nonmagnetic case \cite{dal_corso_density_2007} and will not be discussed further. The other two contributions, providing $\chi_{II'}|_{\psi}$, contain the variation of the KS pseudo-wavefunctions. 
Standard manipulations within DFPT, described in \cite{dal_corso_density-functional_2001,dal_corso_density_2007}, allow to rewrite $\chi_{II'}|_{\psi}$ as
\begin{equation}
    \begin{split}
\chi_{II'}|_{\psi}&=\sum_{\{\sigma\}}\sum_{i,j}\Big[\tilde{\theta}_i-\tilde{\theta}_j\Big]\tilde{\theta}_{ji}\bigg(\big\langle\psi_{j}^{\sigma_3}\big|S^{\sigma_3\sigma_4}\bigg|\frac{\partial\psi_{i}^{\sigma_4}}{\partial\alpha_{I'}}\bigg\rangle\\
&\times\big\langle\psi_{i}^{\sigma_1}\big|P^{I,\sigma\sigma}_{\sigma_1\sigma_2}\big|\psi_{j}^{\sigma_2}\big\rangle\\
&+\big\langle(\mathcal{T}\psi_{j})^{\sigma_3}\big|S^{\sigma_3\sigma_4}\bigg|\bigg(\mathcal{T}\frac{\partial\psi_{i}}{\partial\alpha_{I'}}\bigg)^{\sigma_4}\bigg\rangle\\
&\times\big\langle(\mathcal{T}\psi_{i})^{\sigma_1}\big|\big(\mathcal{T}P^{I}\mathcal{T}^\dagger\big)^{\sigma\sigma}_{\sigma_1\sigma_2}\big|(\mathcal{T}\psi_{j})^{\sigma_2}\big\rangle\bigg)\label{chi}
    \end{split}
\end{equation}
    it should be noted that now, since the $S$ matrix is $\alpha_I$-independent, the change of orthonormalization constraints is just $\sum_{\{\sigma\}}\langle\partial_{I'}\psi_{i}^\sigma|S^{\sigma\sigma'}|\psi_{j}^{\sigma'}\rangle=-\sum_{\{\sigma\}}\langle\psi_{i}^{\sigma}|S^{\sigma\sigma'}|\partial_{I'}\psi_{j}^{\sigma'}\rangle$. Using perturbation theory, the expressions of the first-order wavefunctions are, for the direct one~\cite{dal_corso_density-functional_2001}:
\begin{equation}
\begin{split}
\sum_{\{\sigma\}}\langle\psi_{j}^{\sigma_2}|S^{\sigma_2\sigma_1}&\bigg|\frac{\partial\psi_{i}^{\sigma_1}}{\partial\alpha_{I'}}\bigg\rangle=\sum_{\{\sigma\}}\bigg\langle\psi_{j}^{\sigma_2}\bigg|\frac{\partial_{I'}V^{[\bm{B}]}_{\sigma_2\sigma_1}}{\epsilon_i-\epsilon_j}\bigg|\psi_{i}^{\sigma_1}\bigg\rangle\\
\end{split}
\end{equation}
and for the ``time-reversed" one:
\begin{equation}
\begin{split}
\sum_{\{\sigma\}}\langle(\mathcal{T}\psi_j)^{\sigma_2}|&S^{\sigma_2\sigma_1}\bigg|\bigg(\mathcal{T}\frac{\partial\psi_j}{\partial\alpha_{I'}}\bigg)^{\sigma_1}\bigg\rangle=\\
&\sum_{\{\sigma\}}\bigg\langle(\mathcal{T}\psi_j)^{\sigma_2}\bigg|\frac{\partial_{I'}V^{[-\bm{B}]}_{\sigma_2\sigma_1}}{\epsilon_i-\epsilon_j}\bigg|(\mathcal{T}\psi_i)^{\sigma_1}\bigg\rangle,\label{pertwfcT}
\end{split}
\end{equation}
where the perturbed KS potential has the form \cite{dal_corso_density_2007}
\begin{equation}
\begin{split}
    &\frac{\partial V_{\sigma\sigma'}^{[\bm{B}]}}{\partial\alpha_{I'}}=\sum_{\sigma_1}P^{I,\sigma_1\sigma_1}_{\sigma\sigma'}+\sum_{\sigma_1,\sigma_2}\int d^3r\\
    &\times\bigg(\frac{\partial V_\mathrm{Hxc}(\bm{r})}{\partial\alpha_{I'}}\,\delta_{\sigma_1\sigma_2}
    -\frac{\partial \bm{B}_\mathrm{xc}(\bm{r})}{\partial\alpha_{I'}}\cdot\bm{\sigma}_{\sigma_1\sigma_2}\bigg)\,K_{\sigma\sigma'}^{\sigma_1\sigma_2}(\bm{r});
\end{split}
\end{equation}
here, $V_\mathrm{Hxc}(\bm{r})=\delta E_\mathrm{Hxc}/\delta n(\bm{r})$ and $\bm{B}_\mathrm{xc}(\bm{r})=-\delta E_\mathrm{xc}/\delta \bm{m}(\bm{r})$ are the Hartee exchange-correlation and the magnetic exchange-correlation potentials, respectively. Finally, 
\begin{equation}
\begin{split}
    K_{\sigma\sigma'}^{\sigma_1\sigma_2}&(\bm{r},\bm{r}_1,\bm{r}_2) = \delta(\bm{r}-\bm{r}_1)\delta(\bm{r}-\bm{r}_2)\delta^{\sigma_1\sigma}\delta^{\sigma_2\sigma'}\\
    +&\sum_{I,\nu\nu'}\sum_{\nu_1\nu_2}f^{\sigma\sigma_1}_{\nu\nu_1}Q^{I}_{\nu_1\nu_2}(\bm{r})f^{\sigma_2\sigma'}_{\nu_2\nu'}\beta_{\nu}^I(\bm{r}_2)\,\beta_{\nu'}^I(\bm{r}_1)^\ast
\end{split}
\end{equation}
is the FR US kernel, which is linked to the US $S$ matrix by $S^{\sigma\sigma'}=\sum_{\sigma_1}\int d^3r\,K_{\sigma\sigma'}^{\sigma_1\sigma_1}(\bm{r})$ \cite{dal_corso_density_2007}. The same kernel enters in the definition of the perturbed electronic (spin-resolved) charge density $\partial_I\rho^{\sigma\sigma'}(\bm{r})$, which contains a term similar to Eq. (\ref{chi}), with $P^{I,\sigma\sigma'}_{\sigma_1\sigma_2}$ substituted with $K_{\sigma\sigma'}^{\sigma_1\sigma_2}(\bm{r})$ \cite{dal_corso_density_2007,urru_density_2019}. This shows that the problem has to be solved iteratively until convergence of $\partial_I\rho^{\sigma\sigma'}(\bm{r})$ \cite{baroni_phonons_2001}. Note that in the perturbed KS state of Eq. (\ref{pertwfcT}) the effect of $\mathcal{T}$, when applied to the first-order KS potential, is to reverse the sign of $\bm{B}_\mathrm{xc}$. Collecting together Eqs. (\ref{chi}) and (\ref{pertwfcT}) we obtain the following expression 
\begin{equation}
\begin{split}
    \chi_{II'}|_{\psi}&=\sum_{\{\sigma\}}\sum_{i,j}\frac{\tilde{\theta}_i-\tilde{\theta}_j}{\epsilon_i-\epsilon_j}\,\tilde{\theta}_{ji}\bigg[\bigg\langle\psi_{j}^{\sigma_3}\bigg|\frac{\partial V^{[\bm{B}]}_{\sigma_3\sigma_4}}{\partial\alpha_{I'}}\bigg|\psi_{i}^{\sigma_4}\bigg\rangle\\    &\times\big\langle\psi_{i}^{\sigma_1}\big|P^{I,\sigma\sigma}_{\sigma_1\sigma_2}\big|\psi_{j}^{\sigma_2}\big\rangle\\
    &+\bigg\langle(\mathcal{T}\psi_{j})^{\sigma_3}\bigg|\frac{\partial V^{[-\bm{B}]}_{\sigma_3\sigma_4}}{\partial\alpha_{I'}}\bigg|(\mathcal{T}\psi_i)^{\sigma_4}\bigg\rangle\\
    &\times\big\langle(\mathcal{T}\psi_{i})^{\sigma_1}\big|\big(\mathcal{T}^\dagger P^{I}\mathcal{T}\big)^{\sigma\sigma}_{\sigma_1\sigma_2}\big|(\mathcal{T}\psi_{j})^{\sigma_2}\big\rangle\bigg]\label{chiFinal}\\
    \end{split}
\end{equation}
which has the well-known Lindhard-like form occurring within perturbation theory in the independent-particle approximation \cite{giuliani_quantum_2005}.

As discussed in Ref. \cite{timrov_hubbard_2018}, the former DFPT formulation still relies on the use of supercell, where only the $I$th-atom is perturbed. Hereafter, to denote the $I$th-atom in a supercell we use the notation $\bm{R}_I=\bm{R}_l+\bm{\tau}_\eta$, where $\bm{R}_l$ is the supercell vector and $\bm{\tau}_\eta$ is the basis vector, i.e. $I\rightarrow(l,\eta)$. To recast the perturbative treatment in a form involving only quantities contained in the primitive cell, a monochromatic decomposition of the external potential $P^{I,\sigma\sigma'}_{\sigma_1\sigma_2}=\big(P^I\big)^{\sigma\sigma'}_{\sigma_1\sigma_2}$ is introduced: 
\begin{gather}
P^{I}=\sum_{\bm{q}}^{N_{\bm{q}}}\,\frac{e^{-\iota\bm{R}_l\cdot\bm{q}}}{N_{\bm{q}}}\,P^{\eta}(\bm{q})\label{P::exp}\\
    P^{\eta,\sigma\sigma'}_{\sigma_1\sigma_2}(\bm{q})=\sum_{\bm{k}'}^{N_{\bm{k}}}\sum_m S^{\sigma_1\sigma}\big|\phi^{\eta\sigma}_{m\bm{k}'+\bm{q}}\big\rangle\big\langle \phi^{\eta\sigma'}_{m\bm{k}'}\big|S^{\sigma'\sigma_2}\label{P::q}
\end{gather}
where the Bloch sums 
\begin{equation}
    \begin{split}
        \phi_{m\bm{k}'}^{\eta\sigma}(\bm{r})&=\sum_l \frac{e^{\iota\bm{k}'\cdot\bm{R}_l}}{\sqrt{N_{\bm{k}}}}\,\phi^{l\eta\sigma}_m(\bm{r})=\frac{e^{\iota\bm{k}'\cdot\bm{r}}}{\sqrt{N_{\bm{k}}}}\,v_{m\bm{k}'}^{\eta\sigma}(\bm{r})\label{blochSum}
    \end{split}
\end{equation}
were used \cite{timrov_hubbard_2018} (the localised $\beta$ projectors within Eq.~(\ref{P::q}) undergo a similar decomposition, see Ref.~\cite{timrov_self-consistent_2021}); here $v_{m\bm{k}'}^{\eta\sigma}(\bm{r}+\bm{R}_l)=v_{m\bm{k}'}^{\eta\sigma}(\bm{r})$ is the periodic part, and it makes apparent for the operator $P^{\eta}(\bm{q})$ to enjoy the property \begin{equation*}
\big\langle\bm{r}_1+\bm{R}_l\big|P^{\eta}(\bm{q})\big|\bm{r}_2+\bm{R}_l\big\rangle=e^{\iota\bm{q}\cdot\bm{R}_l}\,\big\langle\bm{r}_1\big|P^{\eta}(\bm{q})\big|\bm{r}_2\big\rangle.
\end{equation*}
Substituting the expansions Eq.~(\ref{P::exp}) and Eq.~(\ref{P::q}) into $\chi_{II'}|_\psi=\chi_{l\eta,l'\eta'}|_\psi$, it can be rewritten as
\begin{equation}
\chi_{l\eta,l'\eta'}|_{\psi}=\sum_{\bm{q}}^{N_{\bm{q}}}\frac{e^{\iota\bm{q}\cdot(\bm{R}_l-\bm{R}_{l'})}}{N_{\bm{q}}}\,\chi_{\eta\eta'}(\bm{q})|_{\psi}
\end{equation}
As routinely done within DFPT \cite{baroni_phonons_2001}, the sum over the empty states (in fact, the one over the $j$ index of Eq. (\ref{chiFinal})) is avoided thanks to the use of Green's function techniques, i.e., by solving a Sternheimer equation where KS projectors over the unoccupied manifold are introduced \cite{baroni_greens-function_1987}. Using the explicit expression of the KS Bloch states $\psi_{n\bm{k}}^\sigma(\bm{r})=\frac{1}{\sqrt{N_{\bm{k}}}}\,e^{\iota\bm{k}\cdot\bm{r}}u_{n\bm{k}}^\sigma(\bm{r})$ (and a transformation similar to Eq. (\ref{blochSum}) for the $\beta$-projectors \cite{urru_density_2019,timrov_self-consistent_2021}), we rewrite everything in terms of the periodic parts $u_{n\bm{k}}^\sigma(\bm{r})$, and the solution is finally recast into the form 
\begin{equation*}
    \begin{split}
        &\chi_{\eta\eta'}(\bm{q})|_{\psi}=\frac{1}{N_{\bm{k}}}\sum_{n\bm{k}}\sum_{\{\sigma\}}\bigg[\big\langle u_{n\bm{k}}^{\sigma_1}\big|P^{\eta,\sigma\sigma}_{\sigma_1\sigma_2}(\bm{q})\big|\Delta_{\bm{q}}u_{n\bm{k}}^{\sigma_2}\big\rangle\\
+&\big\langle(\mathcal{T}u_{n(-\bm{k})})^{\sigma_1}\big|\big(\mathcal{T}P^\eta(\bm{q})\mathcal{T}^\dagger\big)^{\sigma\sigma}_{\sigma_1\sigma_2}\big|(\mathcal{T}\Delta_{(-\bm{q})}u_{n(-\bm{k})})^{\sigma_2}\big\rangle\bigg]
    \end{split}
\end{equation*}
where now the sum over $n\bm{k}$ runs, in fact, only over the occupied states. The two unknowns $\Delta_{\bm{q}} u_{n\bm{k}}^\sigma$ and its ``time-reversed" counterpart  $(\mathcal{T}\Delta_{(-\bm{q})}u_{n(-\bm{k}}))^{\sigma}$ are the solutions of two Sternheimer linear systems: a standard one 
\begin{equation*}
\begin{split}
       \sum_{\sigma'}\Big(H^{[\bm{B}]}_{\sigma\sigma'}- \epsilon_{n\bm{k}}\, &S_{\sigma\sigma'}\Big)\big|\Delta_{\bm{q}}u_{{n\bm{k}}}^{\sigma'}\big\rangle=\\
        &-\sum_{\sigma',\sigma_1}\mathcal{P}^{\dagger\sigma\sigma_1}_{n{\bm{k}+\bm{q}}}\,\frac{\partial V^{[\bm{B}]}_{\sigma_1\sigma'}}{\partial\alpha(\bm{q})}\,\big|u_{{n\bm{k}}}^{\sigma'}\big\rangle
\end{split}
\end{equation*}
and a time-reversed one
\begin{equation*}
\begin{split}
            \sum_{\sigma'}\Big(H^{[-\bm{B}]}_{\sigma\sigma'}&-\epsilon_{n(-\bm{k})}\,S_{\sigma\sigma'}\Big)\big|(\mathcal{T}\Delta_{(-\bm{q})}u_{n(-\bm{k})})^{\sigma'}\big\rangle =\\
            -&\sum_{\sigma_1,\sigma'}\Pi_{n(-\bm{k}-\bm{q})}^{\dagger\sigma\sigma_1}\,\frac{\partial V^{[\bm{-B}]}_{\sigma_1\sigma'}}{\partial\alpha(\bm{q})}\,\big|(\mathcal{T}u_{n(-\bm{k})})^{\sigma'}\big\rangle
\end{split}
\end{equation*}
where we have introduced the (left) projectors: $\mathcal{P}_{n\bm{k}+\bm{q}}^{\dagger\sigma\sigma'}=\tilde{\theta}_{n\bm{k}}-\sum_{m\sigma_1}\beta_{n\bm{k},m\bm{k}+\bm{q}}\,S^{\sigma\sigma_1}\big|\psi_{m\bm{k}+\bm{q}}^{\sigma_1}\big\rangle\big\langle\psi_{m\bm{k}+\bm{q}}^{\sigma'}\big|$ and the time-reversed ones  $\Pi_{n(-\bm{k}-\bm{q})}^{\dagger\sigma\sigma'}=\big(\mathcal{T}\mathcal{P}^{\dagger}_{n(-\bm{k}-\bm{q})}\mathcal{T}^\dagger\big)^{\sigma\sigma'}$, with the usual definition $\beta_{n\bm{k},m\bm{k}'}\equiv\tilde{\theta}_{n\bm{k}}\tilde{\theta}_{n\bm{k},m\bm{k}'}+\tilde{\theta}_{m\bm{k}'}\tilde{\theta}_{m\bm{k}',n\bm{k}}$ \cite{urru_density_2019, de_gironcoli_lattice_1995}.

\section{applications\label{applications}}
In this section, with the formalism described above, we investigate the binary Europium monochalcogenides series EuX (X = O, S, Se, Te), and the cadmium osmate pyrochlore Cd$_2$Os$_2$O$_7$. All the theory illustrated was implemented in the Quantum ESPRESSO distribution within the PW and HP packages \cite{giannozzi_quantum_2009,giannozzi_advanced_2017,timrov_hp_2022}. 

\subsection{Computational details}  
All the first-principles calculations have been carried out using FR US-PPs from the Pslibrary1.0.0 \cite{dal_corso_pseudopotentials_2014}. We employed the generalized-gradient approximation (GGA) with the PBEsol analytical form \cite{perdew_restoring_2008} and -- for the EuX series -- also the local-density approximation (LDA) parametrized by Perdew and Zunger \cite{perdew_self-interaction_1981} as exchange-correlation functional. All the DFT+$U$ calculations have been performed using orthogonalized atomic orbitals, as directly read from the pseudopotential file. The computational details adopted for the materials-specific calculations are reported below; lattice parameters and structural data have been taken from Materials Project \cite{jain_commentary_2013}. High-symmetry paths within the Brillouin zone have been obtained using SeeK-path~\cite{hinuma_band_2017}. Structural analysis of the crystal distortions have been carried out with ISOTROPY(ISOCIF) Software Suite~\cite{campbell_isodisplace_2006}, and visualized with VESTA~\cite{momma_vesta_2008}. The data used to produce the results of this work are available at the Materials Cloud Archive~\cite{MaterialsCloudArchive2023}.
\paragraph{EuX} Even though US-PPs are optimized to be accurate with substantially reduced kinetic energy (KE) cutoffs compared to NC-PPs, the inclusion of $4d$ and $4f$ states in the valence for Eu -- necessary to reduce at minimum the transferability errors -- entails very high KE cutoffs \cite{dal_corso_pseudopotentials_2014}. We used 140~Ry for the wavefunctions and 1120~Ry on the charge density. The Brillouin zone has been sampled with a 18$\times$18$\times$18 $\bm{k}$-points grid for the plane-wave calculations, with a 4$\times$4$\times$4 $\bm{q}$-points mesh used for the evaluation of the Hubbard parameter $U$ applied on Eu. A Gaussian smearing of 0.007~Ry is used to help converge narrow-gap states.

\paragraph{Cd$_2$Os$_2$O$_7$} Numerical simulations for Cd$_2$Os$_2$O$_7$ have been performed using a KE cutoff of 60~Ry and 480~Ry, respectively, for the KS states and the charge density. We used a 8$\times$8$\times$8 $\bm{k}$-mesh for  Brillouin zone sampling and a 2$\times$2$\times$2 $\bm{q}$-points grid for the calculation of the $U$ interaction parameter. The +$U$ correction was applied to the $5d_{5/2}$ and $5d_{3/2}$ states of Os. We employed Gaussian smearing of $0.007$~Ry. 

\begin{figure*}[]
\centering
\includegraphics[width=17cm]{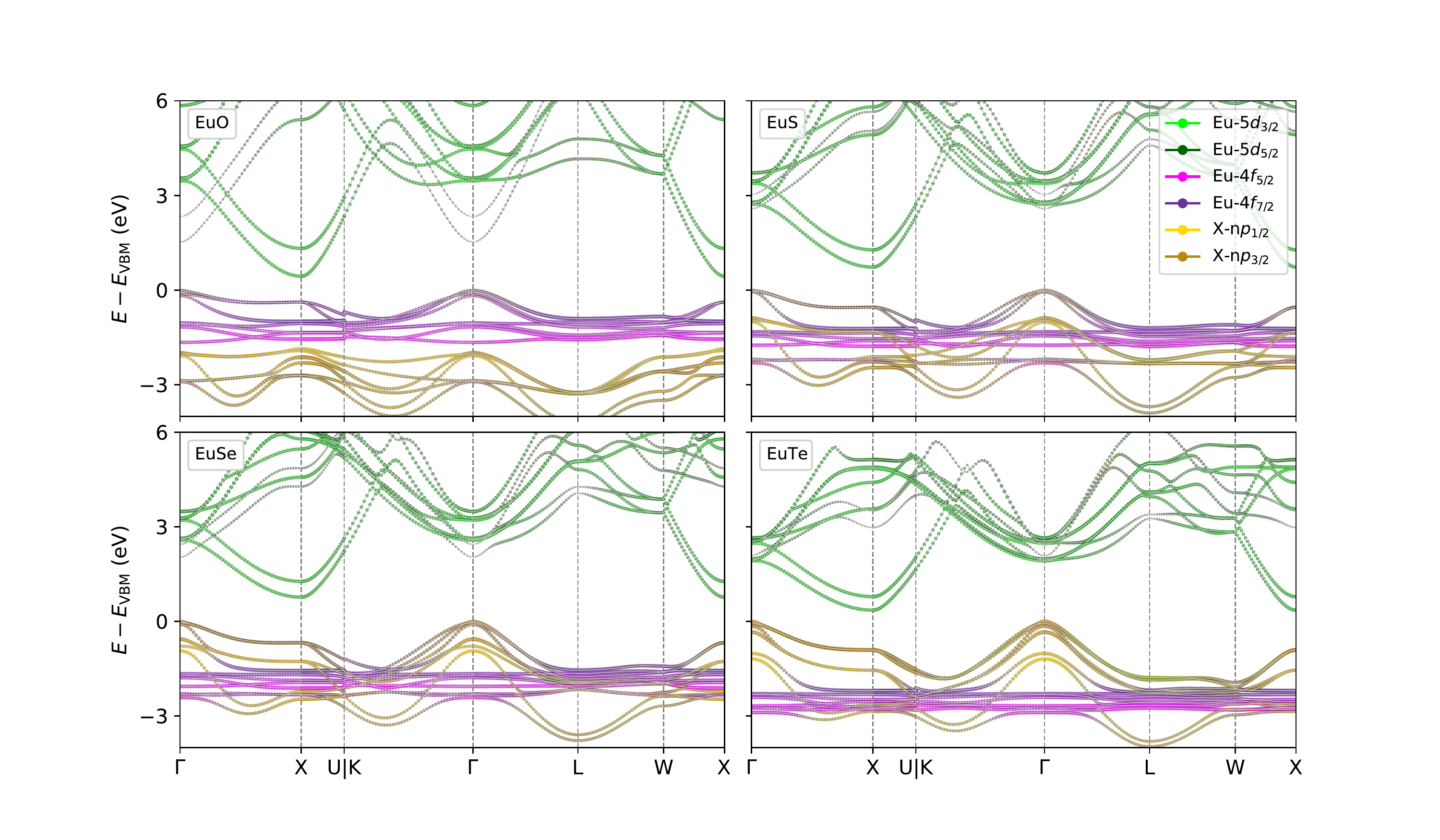}
\caption{Calculated fully-relativistic DFT(GGA--PBEsol)+$U$ band structure of the four EuX compounds considered here (X = O, S, Se, Te) in the $Fm\bar{3}m$ phase, plotted along high-symmetry lines of the Brillouin zone. We use the so-called fatband representation, where the dominant orbital character of each band is also reported. The zero of the energy is at the valence band maximum (VBM).}
\label{fatbands}
\end{figure*} 
\subsection{Europium monoxide and monochalcogenides EuX}
Europium monoxide and monochalcogenides are a series of compounds crystallizing in the relatively simple rock-salt $Fm\Bar{3}m$ space group. In pristine conditions they are insulating materials, and for a Eu$^{2+}$ configuration they have a magnetic moment $m\simeq7$~$\mu_\mathrm{B}$ ($J=7/2$) for half-filled $4f$ electronic states \cite{monteiro_spatially_2013}. EuO and EuS are ferromagnetic semiconductors, EuSe has a complex magnetic phase diagram, eventually becoming an antiferromagnet below 1.8~K, and EuTe develops antiferromagnetism below 9.58~K \cite{mauger_magnetic_1986}. However, both EuSe~\cite{fujiwara_pressure-induced_1982} and EuTe~\cite{ishizuka_pressure-induced_1997} display an antiferromagnetic-to-ferromagnetic phase transition upon application of hydrostatic pressure at low temperatures. The interest in studying these materials arises from the fact that -- besides being easily accessible to study given their structurally simple structure -- EuO and EuS are some of the few naturally occurring ferromagnets at ambient pressure conditions. 

\begin{figure}[]
\centering
\includegraphics[width=8cm]{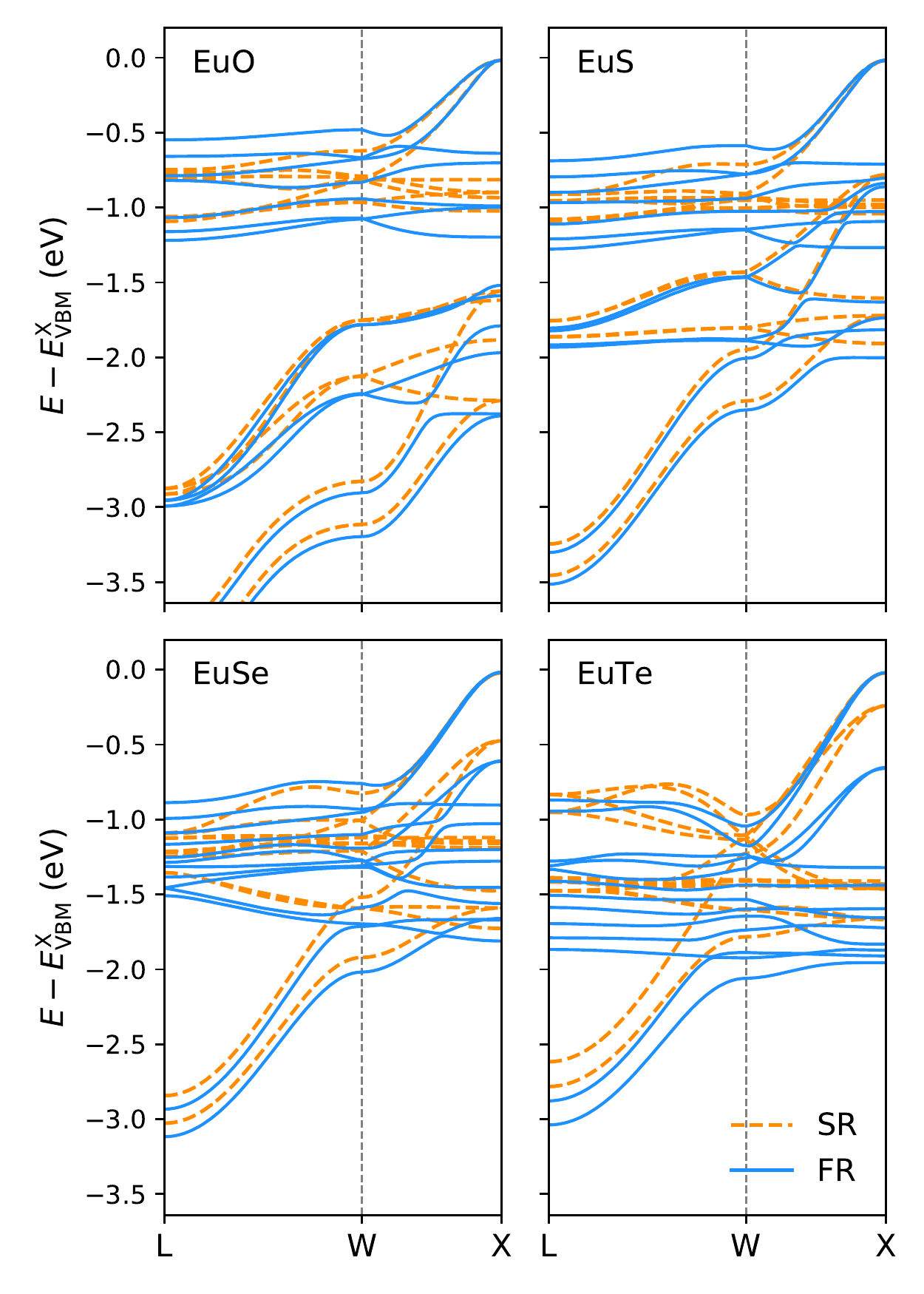}
\caption{Comparison between DFT+$U$ band structures calculated using either scalar-relativistic (SR, dashed orange lines) or fully-relativistic (FR, blue lines) US-PPs along the $\mathrm{L}\rightarrow\mathrm{W}\rightarrow\mathrm{X}$ high symmetry path of the Brillouin zone. The bands are aligned so that the maximum of the valence band at the X point between the SR and the FR electronic structures coincide.\label{bandsSR_vs_FR}}
\end{figure}

We performed spin-polarized \emph{ab initio} calculations in the 2 atoms primitive cell within DFT+$U$ including SOC to the whole EuX series. The $U$ parameter was calculated from first-principles within the scheme described above. Fig. \ref{fatbands} shows the FR GGA+$U$ band structures of the EuX compounds in the $Fm\Bar{3}m$ phase; the orbital character is also highlighted through the so-called fatband representation. At variance with plain DFT, which predicts an erroneous metallic ground states for all the compounds, within DFT+$U$ all the materials are semiconductors. The most relevant aspect in the electronic structure is the progressive deepening in energy of the weakly dispersive Eu--$4f_{7/2}$ and Eu--$4f_{5/2}$ bands as the atomic number of the chalcogen increases, eventually merging with the X--n$p_{3/2}$ and X--n$p_{1/2}$ for X = S, Se and Te; however, this happens without significant $4f$--n$p$ hybridization due to the overall conservation of the $4f$ electronic quantum numbers. This deepening also changes the character of the band gap, which is found between the $4f$ and $4d$ states in EuO and between the n$p$ and $4d$ for the other materials. A comparison between the FR and SR (i.e., without including SOC) band structures is displayed in Fig. \ref{bandsSR_vs_FR}; it shows that for an accurate low-energy electronic description of the Eu--$4f$ states, the inclusion of SOC is essential, due to the considerable broadening of the Eu--$4f$ bandwidth because of spin-orbit splitting.  
\begin{table*}[]
\centering
\caption{Calculated lattice parameter $a_0$ (in \AA), bulk modulus $B_0$ (in GPa), and critical pressure $P_\mathrm{c}$ (in GPa) for the $Fm\bar{3}m$ phase using fully-relativistic (FR) US-PPs and different DFT functionals (LDA and GGA-PBEsol), and compared with experimental data.\label{euoParameters}}
\bgroup
\def\arraystretch{1.05}
\setlength\tabcolsep{0.05in}
\begin{tabular}{lccccccccccccccc}
\toprule
\toprule
\multirow{2}{*}{Method}& \multicolumn{3}{c}{EuO} &&\multicolumn{3}{c}{EuS} & & \multicolumn{3}{c}{EuSe}&&\multicolumn{3}{c}{EuTe}\\
\cline{2-4}\cline{6-8}\cline{10-12}\cline{14-16}
&$a_0$&$B_0$&$P_\mathrm{c}$&&$a_0$&$B_0$&$P_\mathrm{c}$&&$a_0$&$B_0$&$P_\mathrm{c}$&&$a_0$&$B_0$&$P_\mathrm{c}$\\
\midrule
FR--LDA&4.909&125&65&&5.660&67&32&&5.898&57&24&&6.305&43&16\\
FR--GGA&4.971&111&67&&5.732&60&31&&5.978&50&23&&6.393&39&15\\
FR--LDA+$U$&5.040&108&39&&5.819&59&19&&6.043&51&16&&6.434&40&12\\
FR--GGA+$U$&5.102&96&42&&5.893&53&18&&6.123&46&16&&6.523&36&12\\
exp&5.083\footnote{\label{note1} Ref. \cite{jain_commentary_2013}}&110$\pm$5\footnote{\label{note2} Ref. \cite{jayaraman_pressure-volume_1974}}&40\footref{note2}&&5.845\footref{note1}&61$\pm$5\footref{note2}&22\footref{note2}&&6.147\footref{note1}&52$\pm$5\footref{note2}&15\footref{note2}&&6.544\footref{note1}&40$\pm$5\footref{note2}&11\footref{note2}\\
\bottomrule
\bottomrule
\end{tabular}
\egroup
\end{table*}
The calculated magnetic moments are $m^\mathrm{Eu}=6.97,\,6.98\,,6.98$ and $6.99$~$\mu_\mathrm{B}$, for X = O, S, Se and Te respectively, fairly agreeing with the nominally expected ones of 7~$\mu_\mathrm{B}$; this reflects the 7 fully-occupied majority Eu--$4f$ flat bands located right below the Fermi energy. The minority (empty) spin states are instead found significantly higher in energy than the window reported in Fig. \ref{fatbands}, consistently with previous studies~\cite{barbagallo_experimental_2010}. 

It is experimentally known that the EuX compounds exhibit a structural phase transition upon increase of hydrostatic pressure, changing the crystal from the NaCl, fcc-like, $Fm\Bar{3}m$ space group to the CsCl, bcc-like, $Pm\Bar{3}m$ space group \cite{jayaraman_pressure-volume_1974}. We determined the critical pressure of the phase transition $P_\mathrm{c}$ by calculating the equation of state as a function of volumes $V$ for the two phases NaCl-like and CsCl-like, then evaluating the pressure $P$ and finally obtaining the enthaly $\mathcal{H}(P)=E(P)+PV(P)$; the crossing of the two enthalpies $\mathcal{H}_\mathrm{NaCl}$ and $\mathcal{H}_\mathrm{CsCl}$ yields $P_\mathrm{c}$. In Table \ref{euoParameters} the calculated lattice parameter $a_0$, the bulk modulus $B_0$ (evaluated by fitting the 3rd-order Birch-Murnaghan equation of state) and the critical pressure $P_\mathrm{c}$ are reported. For both the $Fm\bar{3}m$ and $Pm\bar{3}m$ phases and the two LDA and GGA functionals, we recalculated the corresponding Hubbard parameters $U$, which are reported in Table \ref{HubbardPara}. In general the best results for $a_0$ are given by the +$U$ correction on top of the GGA-PBEsol funtional. This is not surprising, since the PBEsol functional was developed with the aim of improving equilibrium volumes with respect to LDA and PBE. For $B_0$, the LDA functional provides better results compared with PBEsol, even though the differences are quite small. This trend, together with the level of precision of the results, is also in good agreement with a previous theoretical study employing an impurity solver in the Hubbard I approximation \cite{wan_mechanism_2011}. The critical pressure $P_\mathrm{c}$ calculated with plain DFT functionals instead is systematically overestimated -- up to 65\% with respect to the experimental value for EuO -- while the results obtained with the FR DFT+$U$ functional are significantly more accurate.   

\begin{table}[]
\centering
\caption{Calculated Hubbard $U$ interaction parameters (in eV) using different DFT functionals, for the two NaCl- and CsCl-like phases investigated at the lattice parameters $a_0^\mathrm{NaCl}$ (as reported in Table \ref{euoParameters}) and $a_0^\mathrm{CsCl}=(3.315,3.556,3.776,4.040)$ $\mathrm{\AA}$ for (EuO, EuS, EuSe, EuTe) respectively, using scalar-relativistic (SR) or fully-relativistic (FR) US-PPs.\label{HubbardPara}}
\bgroup
\def\arraystretch{1.1}
\setlength\tabcolsep{0.15in}
\begin{tabular}{lcccc}
\toprule
\toprule
\multirow{2}{*}{PP--funct.}&\multicolumn{4}{c}{$U$ (eV)}\\
\cline{2-5}
&EuO&EuS&EuSe&EuTe\\
\midrule
\multicolumn{5}{c}{$Fm\bar{3}m$ phase}\\
FR--LDA&7.69&7.46&7.55&7.51\\
FR--GGA&7.65&7.65&7.74&7.69\\
SR--GGA&7.87&7.79&7.85&7.79\\
\midrule
\multicolumn{5}{c}{$Pm\bar{3}m$ phase}\\
FR--LDA&7.15&6.98&7.11&7.14\\
FR--GGA&7.10&6.94&7.08&7.11\\
\bottomrule
\bottomrule
\end{tabular}
\egroup
\end{table}

Interestingly, our calculated values of $U$ (reported in Table \ref{euoParameters}) agree well with the empirically-chosen $U$'s (between 6--9~eV), which provide a faithful description of the magnetic exchange interaction parameters in EuX \cite{wan_mechanism_2011,kunes_exchange_2005}. The computed $U$'s are found to be mostly dependent on the chemical environment (i.e. on the crystal phase), while they do not vary appreciably by changing exchange-correlation base functional. It should be noted that the $U$ values obtained with SR- or FR-PP's are very similar, this seems to suggest a good level of transferability among them, provided that the atomic orbitals are very similar (i.e. the PP's are of the same type, in this case US-PPs, and generated through the same code, \texttt{ld1.x} in our case).    

\subsection{Cd$_{\boldsymbol{2}}$Os$_{\boldsymbol{2}}$O$_{\boldsymbol{7}}$}
In recent years magnetic pyrochlore oxides have attracted a lot of interest because of the very rich physics they have shown to host \cite{gardner_magnetic_2010}: from the experimentally-observed spin-ice phases in Ho$_2$Ti$_2$O$_7$ \cite{harris_geometrical_1997} and Dy$_2$Ti$_2$O$_7$ \cite{ramirez_zero-point_1999} to the theoretically proposed three-dimensional topological semimetal in the rare-earth iridates $R_2$Ir$_2$O$_7$ \cite{wan_topological_2011}. Despite the vast number of magnetically disordered frustrated phases appearing in these systems, long-range order was also observed, for example in Cd$_2$Os$_2$O$_7$ \cite{calder_spin-orbit-driven_2016}. At high temperature this material is metallic, but develops a metal-insulator transition when lowering the temperature, with an onset that starts at $T=227$~K and continuously proceeds with a decrease of the Drude response by the free carriers, until a band gap is formed at $T=150$~K \cite{sohn_optical_2015}. The insulating transition is also concomitant with the appearance of tetrahedral antiferromagntic ordering, as confirmed by X-ray diffraction studies \cite{mandrus_continuous_2001,yamaura_tetrahedral_2012}.

\begin{figure}[]
\centering
\includegraphics[width=8.7cm]{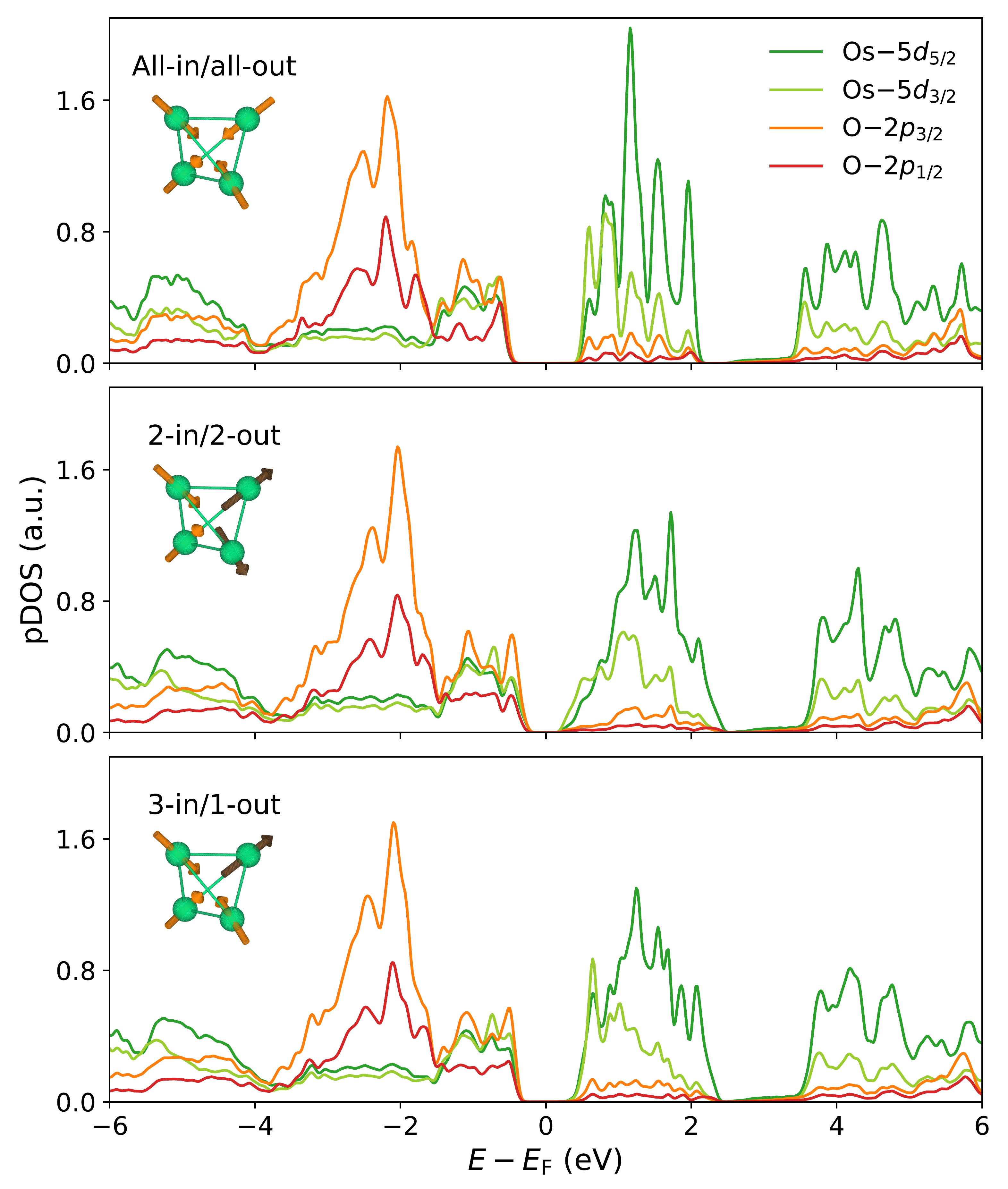}
\caption{Calculated projected density of states (pDOS) for the three magnetic orderings all-in/all-out (AI/AO), 3-in/1-out (3I/1O) and 2-in/2-out (2I/2O). In the figure, the orbital character of the Os--$5d_{5/2}$ and Os--$5d_{3/2}$ is reported, together with the one pertaining to the O--$2p_{3/2}$ and O--$2p_{1/2}$ states. The insets schematically represent the tetrahedral network formed by 4 osmium atoms, with the arrow displaying the direction of the magnetic moment: orange--in and brown--out.}
\label{cdoso}
\end{figure}

Given the tetrahedral network formed by the Os sublattice, the pyrochlore crystal structure allows different noncollinear magnetic orderings to develop. We analyzed three different types of noncollinear magnetic patterns: all-in/all-out (AI/AO), 3-in/1-out (3I/1O)  and 2-in/2-out (2I/2O) configurations; they are depicted in the insets of Fig. \ref{cdoso}. In the first case, the magnetic moments carried by the Os atoms point inward the Os tetrahedron; in the second, 3 magnetic dipoles point inward and 1 outward; in the third, 2 point in and 2 point out. The latter is also the same pattern appearing in the spin-ice state, although in a frustrated (disordered) manner. 

Without the inclusion the Hubbard $E_U$ correction, regardless the initial spin configuration, all the Os ions are nonmagnetic and the electronic ground state converges to an erroneous metallic state. We evaluated the Hubbard $U$ parameter in the FR US-PP scheme illustrated above, and then performed GGA(PBEsol)+$U$ calculations. The inclusion of the Hubbard correction successfully stabilizes all three AI/AO, 3I/1O and 2I/2O magnetic and insulating states. The calculated $U_\mathrm{Os}=4.48$~eV, including the effect of SOC, is considerably larger than the empirical range 0.8--2~eV used in previous \textit{ab initio} studies \cite{sohn_optical_2015,kim_spin-orbit_2020}; therefore, the resulting bad gap is expectedly larger, but the magnetic moments $|\bm{m}_\mathrm{Os}|\simeq1.09$--1.02~$\mu_\mathrm{B}$ in the local $\langle111\rangle$ axis of the Os atoms are similar to those seen in previous simulations \cite{shinaoka_noncollinear_2012}. Also for this material we compared $U_\mathrm{Os}$ obtained with SOC and the one without SOC, at the SR level; in this latter case $U^\mathrm{SR}_\mathrm{Os}=4.61$~eV, which is again very similar to the $U^\mathrm{FR}_\mathrm{Os}=4.48$~eV obtained using FR-PPs.  

\begin{table}[]
\centering
\caption{Simulated types of magnetism: All-in/all-out (AI/AO), 2-in/2-out (2I/2O) and 3-in/1-out (3I/1O), together with the absolute value of the magnetic moments on the Os ions $|\bm{m}_\mathrm{Os}|$, energy difference $\Delta E=E-E_\mathrm{AI/AO}$ with respect to the AI/AO configuration, and space group of the relaxed structure.\label{tab:cd2os2o7}}
\bgroup
\def\arraystretch{1.1}
\setlength\tabcolsep{0.11in}
\begin{tabular}{cccc}
\toprule
\toprule
magnetism&$m_\mathrm{Os}$~($\mu_\mathrm{B}$)&$\Delta E$~(meV/f.u.)&\makecell{Space \\group}\\
\midrule
\midrule
\multicolumn{4}{c}{exp}\\
AI/AO\footnote{\label{note-a}Ref. \cite{yamaura_tetrahedral_2012}}&&&$Fd$--$3m$\footnote{\label{note_b}Ref. \cite{jain_commentary_2013}}\\
\midrule
\multicolumn{4}{c}{$\mathrm{GGA(PBEsol)}+U=4.48$~eV}\\
AI/AO&1.09&0&$Fd$--$3m$\\
2I/2O&1.09&+85&$I4_1/amd$\\
3I/1O&1.03(Os$_1$)&+57&$R$--$3m$\\
&1.10(Os$_2$)&&\\
\bottomrule
\bottomrule
\end{tabular}
\egroup
\end{table}
Fig. \ref{cdoso} shows the calculated GGA+$U$ projected density of states (pDOS) for the three magnetic states; each structure has relaxed lattice parameters and atomic positions. The different magnetic state has a marginal effect on the electronic structure, changing mostly the Os--$5d_{5/2}$ and the Os--$5d_{3/2}$ conduction states and slightly the amplitude of the band gap. Nevertheless, different forms of magnetism can still produce qualitative changes in the crystal structure. As shown in Table \ref{tab:cd2os2o7}, the AI/AO state mantains the cubic space group $Fd$--$3m$ (227), consistently with experiments, where there are two inequivalent oxygen atoms O$_1$(in the $8b$ Wyckoff position) and O$_2$($48f$), the latter surrounding all the equivalent Os($16c$). On the contrary, the 2I/2O and the 3I/1O orderings change the original space group. For the first, the material acquires a tetragonal $I4_1/amd$ (141) distortion, containing three inequivalent O: O$_1$($4a$), O$_2$($16g$) and O$_3$($8e$), where always four O$_2$ and two O$_3$ surround the Os$(8d)$ atoms, which are again all equivalent. In the 3I/1O ordering, the relaxed space group is the rhombohedral $R$--$3m$ (166); in this case there are two inequivalent osmium atoms: Os$_1$($3a$, the 1-out), surrounded by six O$_2$($18h$), and Os$_2$($9d$ the 3-in), enclosed within four O$_3$($18h$) and two O$_2$. This is reflected in a slightly different magnetic moment on Os$_1$, as reported in Table \ref{tab:cd2os2o7}. By analyzing the electronic energetics of the three investigated magnetic orderings, shown in Table \ref{tab:cd2os2o7}, we find the AI/AO state to be the ground state of the material, in accordance with experiments \cite{mandrus_continuous_2001,yamaura_tetrahedral_2012}. 

\section{summary and conclusions}
We have presented a comprehensive \emph{ab initio} scheme to perform first-principles simulations in presence of both: ($i$) noncollinear magnetism and/or strong spin-orbit coupling requiring a fully-relativistic treatment, and ($ii$) significant electronic localization requiring Hubbard corrections to local/semi-local DFT functionals. The formalism is extended to a modern and widespread pseudopotential approach, the ultrasoft scheme \cite{vanderbilt_soft_1990}, which allows accurate and precise simulation on $d$ and $f$ electrons preserving modest kinetic energy cutoffs in a plane-wave expansion of the pseudo-wavefunctions. We also developed a density-functional perturbation theory approach to the calculation of the Hubbard interaction parameters $U$ in case of noncollinear magnetism and/or spin-orbit coupling. In the application of the formulation to real materials, it has been shown to be able to reproduce both qualitatively and quantitatively the known experimental features of the systems studied. For EuX, the equilibrium volume, the bulk modulus and the critical pressure $P_\mathrm{c}$ are accurately reproduced; significantly, for $P_\mathrm{c}$ the effects of the $+U$ correction are larger when $P_\mathrm{c}$ from pure DFT deviates more from the experimental values (e.g. for EuO), while they are smaller when DFT is closer to experiments (e.g. for EuTe). For Cd$_2$Os$_2$O$_7$, we determined the energetics of different magnetic noncollinear orderings and assessed their influence within the variable-cell structural relaxation process. It is found that different magnetic orderings lower their energy by distorting into crystal structures with lower symmetries. However, at the GGA--PBEsol+$U$ level of theory, the all-in/all-out (AI/AO) magnetic ordering yields the lower energy state, in a relaxed crystal structure displaying the $Fd$--$3m$ space group, in both cases consistently with experiments.       

The \emph{ab initio} scheme we developed, allowed us to incorporate for the first time within the linear-response method, the effect of spin-orbit coupling (SOC) and noncollinear magnetism in the determination of the Hubbard $U$ parameters. In general, the inclusion of such effects for the evaluation of $U$ is not a trivial task: for example, in the constrained random-phase approximation (cRPA) approach~\cite{aryasetiawan_frequency-dependent_2004}, the inclusion of SOC in the calculation of the $U$ parameters can become technically difficult because of the involvement of complex-valued Wannier spinors in the \textit{ab initio} treatment~\cite{liu_comparative_2020}. From our results, the very close similarity between the Hubbard $U$ values calculated with and without SOC (differing, at maximum, less than 3\% for Cd$_2$Os$_2$O$_7$) seems to suggest that -- within the linear-response approach -- a fully-relativistic DFT+$U$ calculation carried out with Hubbard parameters obtained from a scalar-relativistic pseudopotential (in the collinear approximation) constitutes a reasonable approximation to the complete fully-relativistic computational setup, provided that the scalar-relativistic and the fully-relativistic pseudopotentials are closely matched (e.g. generated with the same parameters and with the same code). 
However, further extensive tests are required to further corroborate this hypothesis.

\section{acknowledgements}
We gratefully acknowledge M. Cococcioni, I. Timrov and A. Urru for useful discussions, and A. Dal Corso for carefully reading the manuscript. This research was supported by the NCCR MARVEL, a National Centre of Competence in Research, funded by the Swiss National Science Foundation (grant number 182892). Computer time was provided by CSCS (Piz Daint) through project No.~s1073.
\bibliography{references,matCloudRef}

\begin{thebibliography}{64}%
\makeatletter
\providecommand \@ifxundefined [1]{%
 \@ifx{#1\undefined}
}%
\providecommand \@ifnum [1]{%
 \ifnum #1\expandafter \@firstoftwo
 \else \expandafter \@secondoftwo
 \fi
}%
\providecommand \@ifx [1]{%
 \ifx #1\expandafter \@firstoftwo
 \else \expandafter \@secondoftwo
 \fi
}%
\providecommand \natexlab [1]{#1}%
\providecommand \enquote  [1]{``#1''}%
\providecommand \bibnamefont  [1]{#1}%
\providecommand \bibfnamefont [1]{#1}%
\providecommand \citenamefont [1]{#1}%
\providecommand \href@noop [0]{\@secondoftwo}%
\providecommand \href [0]{\begingroup \@sanitize@url \@href}%
\providecommand \@href[1]{\@@startlink{#1}\@@href}%
\providecommand \@@href[1]{\endgroup#1\@@endlink}%
\providecommand \@sanitize@url [0]{\catcode `\\12\catcode `\$12\catcode
  `\&12\catcode `\#12\catcode `\^12\catcode `\_12\catcode `\%12\relax}%
\providecommand \@@startlink[1]{}%
\providecommand \@@endlink[0]{}%
\providecommand \url  [0]{\begingroup\@sanitize@url \@url }%
\providecommand \@url [1]{\endgroup\@href {#1}{\urlprefix }}%
\providecommand \urlprefix  [0]{URL }%
\providecommand \Eprint [0]{\href }%
\providecommand \doibase [0]{http://dx.doi.org/}%
\providecommand \selectlanguage [0]{\@gobble}%
\providecommand \bibinfo  [0]{\@secondoftwo}%
\providecommand \bibfield  [0]{\@secondoftwo}%
\providecommand \translation [1]{[#1]}%
\providecommand \BibitemOpen [0]{}%
\providecommand \bibitemStop [0]{}%
\providecommand \bibitemNoStop [0]{.\EOS\space}%
\providecommand \EOS [0]{\spacefactor3000\relax}%
\providecommand \BibitemShut  [1]{\csname bibitem#1\endcsname}%
\let\auto@bib@innerbib\@empty
\bibitem [{\citenamefont {Perdew}\ and\ \citenamefont
  {Zunger}(1981)}]{perdew_self-interaction_1981}%
  \BibitemOpen
  \bibfield  {author} {\bibinfo {author} {\bibfnamefont {J.~P.}\ \bibnamefont
  {Perdew}}\ and\ \bibinfo {author} {\bibfnamefont {A.}~\bibnamefont
  {Zunger}},\ }\href {\doibase 10.1103/PhysRevB.23.5048} {\bibfield  {journal}
  {\bibinfo  {journal} {Physical Review B}\ }\textbf {\bibinfo {volume} {23}},\
  \bibinfo {pages} {5048} (\bibinfo {year} {1981})}\BibitemShut {NoStop}%
\bibitem [{\citenamefont {Anisimov}\ \emph {et~al.}(1991)\citenamefont
  {Anisimov}, \citenamefont {Zaanen},\ and\ \citenamefont
  {Andersen}}]{anisimov_band_1991}%
  \BibitemOpen
  \bibfield  {author} {\bibinfo {author} {\bibfnamefont {V.~I.}\ \bibnamefont
  {Anisimov}}, \bibinfo {author} {\bibfnamefont {J.}~\bibnamefont {Zaanen}}, \
  and\ \bibinfo {author} {\bibfnamefont {O.~K.}\ \bibnamefont {Andersen}},\
  }\href {\doibase 10.1103/PhysRevB.44.943} {\bibfield  {journal} {\bibinfo
  {journal} {Physical Review B}\ }\textbf {\bibinfo {volume} {44}},\ \bibinfo
  {pages} {943} (\bibinfo {year} {1991})}\BibitemShut {NoStop}%
\bibitem [{\citenamefont {Anisimov}\ \emph {et~al.}(1997)\citenamefont
  {Anisimov}, \citenamefont {Aryasetiawan},\ and\ \citenamefont
  {Lichtenstein}}]{anisimov_first-principles_1997}%
  \BibitemOpen
  \bibfield  {author} {\bibinfo {author} {\bibfnamefont {V.~I.}\ \bibnamefont
  {Anisimov}}, \bibinfo {author} {\bibfnamefont {F.}~\bibnamefont
  {Aryasetiawan}}, \ and\ \bibinfo {author} {\bibfnamefont {A.~I.}\
  \bibnamefont {Lichtenstein}},\ }\href {\doibase 10.1088/0953-8984/9/4/002}
  {\bibfield  {journal} {\bibinfo  {journal} {Journal of Physics: Condensed
  Matter}\ }\textbf {\bibinfo {volume} {9}},\ \bibinfo {pages} {767} (\bibinfo
  {year} {1997})}\BibitemShut {NoStop}%
\bibitem [{\citenamefont {Himmetoglu}\ \emph {et~al.}(2014)\citenamefont
  {Himmetoglu}, \citenamefont {Floris}, \citenamefont {de~Gironcoli},\ and\
  \citenamefont {Cococcioni}}]{himmetoglu_hubbard-corrected_2014}%
  \BibitemOpen
  \bibfield  {author} {\bibinfo {author} {\bibfnamefont {B.}~\bibnamefont
  {Himmetoglu}}, \bibinfo {author} {\bibfnamefont {A.}~\bibnamefont {Floris}},
  \bibinfo {author} {\bibfnamefont {S.}~\bibnamefont {de~Gironcoli}}, \ and\
  \bibinfo {author} {\bibfnamefont {M.}~\bibnamefont {Cococcioni}},\ }\href
  {\doibase 10.1002/qua.24521} {\bibfield  {journal} {\bibinfo  {journal}
  {International Journal of Quantum Chemistry}\ }\textbf {\bibinfo {volume}
  {114}},\ \bibinfo {pages} {14} (\bibinfo {year} {2014})}\BibitemShut
  {NoStop}%
\bibitem [{\citenamefont {Kulik}\ \emph {et~al.}(2006)\citenamefont {Kulik},
  \citenamefont {Cococcioni}, \citenamefont {Scherlis},\ and\ \citenamefont
  {Marzari}}]{kulik_density_2006}%
  \BibitemOpen
  \bibfield  {author} {\bibinfo {author} {\bibfnamefont {H.~J.}\ \bibnamefont
  {Kulik}}, \bibinfo {author} {\bibfnamefont {M.}~\bibnamefont {Cococcioni}},
  \bibinfo {author} {\bibfnamefont {D.~A.}\ \bibnamefont {Scherlis}}, \ and\
  \bibinfo {author} {\bibfnamefont {N.}~\bibnamefont {Marzari}},\ }\href
  {\doibase 10.1103/PhysRevLett.97.103001} {\bibfield  {journal} {\bibinfo
  {journal} {Physical Review Letters}\ }\textbf {\bibinfo {volume} {97}},\
  \bibinfo {pages} {103001} (\bibinfo {year} {2006})}\BibitemShut {NoStop}%
\bibitem [{\citenamefont {Anisimov}\ \emph {et~al.}(1993)\citenamefont
  {Anisimov}, \citenamefont {Solovyev}, \citenamefont {Korotin}, \citenamefont
  {Czyżyk},\ and\ \citenamefont
  {Sawatzky}}]{anisimov_density-functional_1993}%
  \BibitemOpen
  \bibfield  {author} {\bibinfo {author} {\bibfnamefont {V.~I.}\ \bibnamefont
  {Anisimov}}, \bibinfo {author} {\bibfnamefont {I.~V.}\ \bibnamefont
  {Solovyev}}, \bibinfo {author} {\bibfnamefont {M.~A.}\ \bibnamefont
  {Korotin}}, \bibinfo {author} {\bibfnamefont {M.~T.}\ \bibnamefont
  {Czyżyk}}, \ and\ \bibinfo {author} {\bibfnamefont {G.~A.}\ \bibnamefont
  {Sawatzky}},\ }\href {\doibase 10.1103/PhysRevB.48.16929} {\bibfield
  {journal} {\bibinfo  {journal} {Physical Review B}\ }\textbf {\bibinfo
  {volume} {48}},\ \bibinfo {pages} {16929} (\bibinfo {year}
  {1993})}\BibitemShut {NoStop}%
\bibitem [{\citenamefont {Cococcioni}\ and\ \citenamefont
  {de~Gironcoli}(2005)}]{cococcioni_linear_2005}%
  \BibitemOpen
  \bibfield  {author} {\bibinfo {author} {\bibfnamefont {M.}~\bibnamefont
  {Cococcioni}}\ and\ \bibinfo {author} {\bibfnamefont {S.}~\bibnamefont
  {de~Gironcoli}},\ }\href {\doibase 10.1103/PhysRevB.71.035105} {\bibfield
  {journal} {\bibinfo  {journal} {Physical Review B}\ }\textbf {\bibinfo
  {volume} {71}},\ \bibinfo {pages} {035105} (\bibinfo {year}
  {2005})}\BibitemShut {NoStop}%
\bibitem [{\citenamefont {Agapito}\ \emph {et~al.}(2015)\citenamefont
  {Agapito}, \citenamefont {Curtarolo},\ and\ \citenamefont
  {Buongiorno~Nardelli}}]{agapito_reformulation_2015}%
  \BibitemOpen
  \bibfield  {author} {\bibinfo {author} {\bibfnamefont {L.~A.}\ \bibnamefont
  {Agapito}}, \bibinfo {author} {\bibfnamefont {S.}~\bibnamefont {Curtarolo}},
  \ and\ \bibinfo {author} {\bibfnamefont {M.}~\bibnamefont
  {Buongiorno~Nardelli}},\ }\href {\doibase 10.1103/PhysRevX.5.011006}
  {\bibfield  {journal} {\bibinfo  {journal} {Physical Review X}\ }\textbf
  {\bibinfo {volume} {5}},\ \bibinfo {pages} {011006} (\bibinfo {year}
  {2015})}\BibitemShut {NoStop}%
\bibitem [{\citenamefont {Kulik}\ and\ \citenamefont
  {Marzari}(2008)}]{kulik_self-consistent_2008}%
  \BibitemOpen
  \bibfield  {author} {\bibinfo {author} {\bibfnamefont {H.~J.}\ \bibnamefont
  {Kulik}}\ and\ \bibinfo {author} {\bibfnamefont {N.}~\bibnamefont
  {Marzari}},\ }\href {\doibase 10.1063/1.2987444} {\bibfield  {journal}
  {\bibinfo  {journal} {The Journal of Chemical Physics}\ }\textbf {\bibinfo
  {volume} {129}},\ \bibinfo {pages} {134314} (\bibinfo {year}
  {2008})}\BibitemShut {NoStop}%
\bibitem [{\citenamefont {Pyykko}(1988)}]{pyykko_relativistic_1988}%
  \BibitemOpen
  \bibfield  {author} {\bibinfo {author} {\bibfnamefont {P.}~\bibnamefont
  {Pyykko}},\ }\href {\doibase 10.1021/cr00085a006} {\bibfield  {journal}
  {\bibinfo  {journal} {Chemical Reviews}\ }\textbf {\bibinfo {volume} {88}},\
  \bibinfo {pages} {563} (\bibinfo {year} {1988})}\BibitemShut {NoStop}%
\bibitem [{\citenamefont {Kim}\ \emph {et~al.}(2008)\citenamefont {Kim},
  \citenamefont {Jin}, \citenamefont {Moon}, \citenamefont {Kim}, \citenamefont
  {Park}, \citenamefont {Leem}, \citenamefont {Yu}, \citenamefont {Noh},
  \citenamefont {Kim}, \citenamefont {Oh}, \citenamefont {Park}, \citenamefont
  {Durairaj}, \citenamefont {Cao},\ and\ \citenamefont
  {Rotenberg}}]{kim_novel_2008}%
  \BibitemOpen
  \bibfield  {author} {\bibinfo {author} {\bibfnamefont {B.~J.}\ \bibnamefont
  {Kim}}, \bibinfo {author} {\bibfnamefont {H.}~\bibnamefont {Jin}}, \bibinfo
  {author} {\bibfnamefont {S.~J.}\ \bibnamefont {Moon}}, \bibinfo {author}
  {\bibfnamefont {J.-Y.}\ \bibnamefont {Kim}}, \bibinfo {author} {\bibfnamefont
  {B.-G.}\ \bibnamefont {Park}}, \bibinfo {author} {\bibfnamefont {C.~S.}\
  \bibnamefont {Leem}}, \bibinfo {author} {\bibfnamefont {J.}~\bibnamefont
  {Yu}}, \bibinfo {author} {\bibfnamefont {T.~W.}\ \bibnamefont {Noh}},
  \bibinfo {author} {\bibfnamefont {C.}~\bibnamefont {Kim}}, \bibinfo {author}
  {\bibfnamefont {S.-J.}\ \bibnamefont {Oh}}, \bibinfo {author} {\bibfnamefont
  {J.-H.}\ \bibnamefont {Park}}, \bibinfo {author} {\bibfnamefont
  {V.}~\bibnamefont {Durairaj}}, \bibinfo {author} {\bibfnamefont
  {G.}~\bibnamefont {Cao}}, \ and\ \bibinfo {author} {\bibfnamefont
  {E.}~\bibnamefont {Rotenberg}},\ }\href {\doibase
  10.1103/PhysRevLett.101.076402} {\bibfield  {journal} {\bibinfo  {journal}
  {Physical Review Letters}\ }\textbf {\bibinfo {volume} {101}},\ \bibinfo
  {pages} {076402} (\bibinfo {year} {2008})}\BibitemShut {NoStop}%
\bibitem [{\citenamefont {Wan}\ \emph {et~al.}(2011{\natexlab{a}})\citenamefont
  {Wan}, \citenamefont {Turner}, \citenamefont {Vishwanath},\ and\
  \citenamefont {Savrasov}}]{wan_topological_2011}%
  \BibitemOpen
  \bibfield  {author} {\bibinfo {author} {\bibfnamefont {X.}~\bibnamefont
  {Wan}}, \bibinfo {author} {\bibfnamefont {A.~M.}\ \bibnamefont {Turner}},
  \bibinfo {author} {\bibfnamefont {A.}~\bibnamefont {Vishwanath}}, \ and\
  \bibinfo {author} {\bibfnamefont {S.~Y.}\ \bibnamefont {Savrasov}},\ }\href
  {\doibase 10.1103/PhysRevB.83.205101} {\bibfield  {journal} {\bibinfo
  {journal} {Physical Review B}\ }\textbf {\bibinfo {volume} {83}},\ \bibinfo
  {pages} {205101} (\bibinfo {year} {2011}{\natexlab{a}})}\BibitemShut
  {NoStop}%
\bibitem [{\citenamefont {Zhang}\ \emph {et~al.}(2012)\citenamefont {Zhang},
  \citenamefont {Zhang}, \citenamefont {Wang}, \citenamefont {Felser},\ and\
  \citenamefont {Zhang}}]{zhang_actinide_2012}%
  \BibitemOpen
  \bibfield  {author} {\bibinfo {author} {\bibfnamefont {X.}~\bibnamefont
  {Zhang}}, \bibinfo {author} {\bibfnamefont {H.}~\bibnamefont {Zhang}},
  \bibinfo {author} {\bibfnamefont {J.}~\bibnamefont {Wang}}, \bibinfo {author}
  {\bibfnamefont {C.}~\bibnamefont {Felser}}, \ and\ \bibinfo {author}
  {\bibfnamefont {S.-C.}\ \bibnamefont {Zhang}},\ }\href {\doibase
  10.1126/science.1216184} {\bibfield  {journal} {\bibinfo  {journal}
  {Science}\ }\textbf {\bibinfo {volume} {335}},\ \bibinfo {pages} {1464}
  (\bibinfo {year} {2012})}\BibitemShut {NoStop}%
\bibitem [{\citenamefont {Erickson}\ \emph {et~al.}(2007)\citenamefont
  {Erickson}, \citenamefont {Misra}, \citenamefont {Miller}, \citenamefont
  {Gupta}, \citenamefont {Schlesinger}, \citenamefont {Harrison}, \citenamefont
  {Kim},\ and\ \citenamefont {Fisher}}]{erickson_ferromagnetism_2007}%
  \BibitemOpen
  \bibfield  {author} {\bibinfo {author} {\bibfnamefont {A.~S.}\ \bibnamefont
  {Erickson}}, \bibinfo {author} {\bibfnamefont {S.}~\bibnamefont {Misra}},
  \bibinfo {author} {\bibfnamefont {G.~J.}\ \bibnamefont {Miller}}, \bibinfo
  {author} {\bibfnamefont {R.~R.}\ \bibnamefont {Gupta}}, \bibinfo {author}
  {\bibfnamefont {Z.}~\bibnamefont {Schlesinger}}, \bibinfo {author}
  {\bibfnamefont {W.~A.}\ \bibnamefont {Harrison}}, \bibinfo {author}
  {\bibfnamefont {J.~M.}\ \bibnamefont {Kim}}, \ and\ \bibinfo {author}
  {\bibfnamefont {I.~R.}\ \bibnamefont {Fisher}},\ }\href {\doibase
  10.1103/PhysRevLett.99.016404} {\bibfield  {journal} {\bibinfo  {journal}
  {Physical Review Letters}\ }\textbf {\bibinfo {volume} {99}},\ \bibinfo
  {pages} {016404} (\bibinfo {year} {2007})}\BibitemShut {NoStop}%
\bibitem [{\citenamefont {Dudarev}\ \emph {et~al.}(2019)\citenamefont
  {Dudarev}, \citenamefont {Liu}, \citenamefont {Andersson}, \citenamefont
  {Stanek}, \citenamefont {Ozaki},\ and\ \citenamefont
  {Franchini}}]{dudarev_parametrization_2019}%
  \BibitemOpen
  \bibfield  {author} {\bibinfo {author} {\bibfnamefont {S.~L.}\ \bibnamefont
  {Dudarev}}, \bibinfo {author} {\bibfnamefont {P.}~\bibnamefont {Liu}},
  \bibinfo {author} {\bibfnamefont {D.~A.}\ \bibnamefont {Andersson}}, \bibinfo
  {author} {\bibfnamefont {C.~R.}\ \bibnamefont {Stanek}}, \bibinfo {author}
  {\bibfnamefont {T.}~\bibnamefont {Ozaki}}, \ and\ \bibinfo {author}
  {\bibfnamefont {C.}~\bibnamefont {Franchini}},\ }\href {\doibase
  10.1103/PhysRevMaterials.3.083802} {\bibfield  {journal} {\bibinfo  {journal}
  {Physical Review Materials}\ }\textbf {\bibinfo {volume} {3}},\ \bibinfo
  {pages} {083802} (\bibinfo {year} {2019})}\BibitemShut {NoStop}%
\bibitem [{\citenamefont {Tancogne-Dejean}\ \emph {et~al.}(2017)\citenamefont
  {Tancogne-Dejean}, \citenamefont {Oliveira},\ and\ \citenamefont
  {Rubio}}]{tancogne-dejean_self-consistent_2017}%
  \BibitemOpen
  \bibfield  {author} {\bibinfo {author} {\bibfnamefont {N.}~\bibnamefont
  {Tancogne-Dejean}}, \bibinfo {author} {\bibfnamefont {M.~J.~T.}\ \bibnamefont
  {Oliveira}}, \ and\ \bibinfo {author} {\bibfnamefont {A.}~\bibnamefont
  {Rubio}},\ }\href {\doibase 10.1103/PhysRevB.96.245133} {\bibfield  {journal}
  {\bibinfo  {journal} {Physical Review B}\ }\textbf {\bibinfo {volume} {96}},\
  \bibinfo {pages} {245133} (\bibinfo {year} {2017})}\BibitemShut {NoStop}%
\bibitem [{\citenamefont {Vanderbilt}(1990)}]{vanderbilt_soft_1990}%
  \BibitemOpen
  \bibfield  {author} {\bibinfo {author} {\bibfnamefont {D.}~\bibnamefont
  {Vanderbilt}},\ }\href {\doibase 10.1103/PhysRevB.41.7892} {\bibfield
  {journal} {\bibinfo  {journal} {Physical Review B}\ }\textbf {\bibinfo
  {volume} {41}},\ \bibinfo {pages} {7892} (\bibinfo {year}
  {1990})}\BibitemShut {NoStop}%
\bibitem [{\citenamefont {Corso}\ and\ \citenamefont
  {Conte}(2005)}]{corso_spin-orbit_2005}%
  \BibitemOpen
  \bibfield  {author} {\bibinfo {author} {\bibfnamefont {A.~D.}\ \bibnamefont
  {Corso}}\ and\ \bibinfo {author} {\bibfnamefont {A.~M.}\ \bibnamefont
  {Conte}},\ }\href {\doibase 10.1103/PhysRevB.71.115106} {\bibfield  {journal}
  {\bibinfo  {journal} {Physical Review B}\ }\textbf {\bibinfo {volume} {71}},\
  \bibinfo {pages} {115106} (\bibinfo {year} {2005})}\BibitemShut {NoStop}%
\bibitem [{\citenamefont {Giannozzi}\ \emph {et~al.}(2017)\citenamefont
  {Giannozzi}, \citenamefont {Andreussi}, \citenamefont {Brumme}, \citenamefont
  {Bunau}, \citenamefont {Buongiorno~Nardelli}, \citenamefont {Calandra},
  \citenamefont {Car}, \citenamefont {Cavazzoni}, \citenamefont {Ceresoli},
  \citenamefont {Cococcioni}, \citenamefont {Colonna}, \citenamefont
  {Carnimeo}, \citenamefont {Dal~Corso}, \citenamefont {de~Gironcoli},
  \citenamefont {Delugas}, \citenamefont {DiStasio}, \citenamefont {Ferretti},
  \citenamefont {Floris}, \citenamefont {Fratesi}, \citenamefont {Fugallo},
  \citenamefont {Gebauer}, \citenamefont {Gerstmann}, \citenamefont {Giustino},
  \citenamefont {Gorni}, \citenamefont {Jia}, \citenamefont {Kawamura},
  \citenamefont {Ko}, \citenamefont {Kokalj}, \citenamefont {Küçükbenli},
  \citenamefont {Lazzeri}, \citenamefont {Marsili}, \citenamefont {Marzari},
  \citenamefont {Mauri}, \citenamefont {Nguyen}, \citenamefont {Nguyen},
  \citenamefont {Otero-de-la Roza}, \citenamefont {Paulatto}, \citenamefont
  {Poncé}, \citenamefont {Rocca}, \citenamefont {Sabatini}, \citenamefont
  {Santra}, \citenamefont {Schlipf}, \citenamefont {Seitsonen}, \citenamefont
  {Smogunov}, \citenamefont {Timrov}, \citenamefont {Thonhauser}, \citenamefont
  {Umari}, \citenamefont {Vast}, \citenamefont {Wu},\ and\ \citenamefont
  {Baroni}}]{giannozzi_advanced_2017}%
  \BibitemOpen
  \bibfield  {author} {\bibinfo {author} {\bibfnamefont {P.}~\bibnamefont
  {Giannozzi}}, \bibinfo {author} {\bibfnamefont {O.}~\bibnamefont
  {Andreussi}}, \bibinfo {author} {\bibfnamefont {T.}~\bibnamefont {Brumme}},
  \bibinfo {author} {\bibfnamefont {O.}~\bibnamefont {Bunau}}, \bibinfo
  {author} {\bibfnamefont {M.}~\bibnamefont {Buongiorno~Nardelli}}, \bibinfo
  {author} {\bibfnamefont {M.}~\bibnamefont {Calandra}}, \bibinfo {author}
  {\bibfnamefont {R.}~\bibnamefont {Car}}, \bibinfo {author} {\bibfnamefont
  {C.}~\bibnamefont {Cavazzoni}}, \bibinfo {author} {\bibfnamefont
  {D.}~\bibnamefont {Ceresoli}}, \bibinfo {author} {\bibfnamefont
  {M.}~\bibnamefont {Cococcioni}}, \bibinfo {author} {\bibfnamefont
  {N.}~\bibnamefont {Colonna}}, \bibinfo {author} {\bibfnamefont
  {I.}~\bibnamefont {Carnimeo}}, \bibinfo {author} {\bibfnamefont
  {A.}~\bibnamefont {Dal~Corso}}, \bibinfo {author} {\bibfnamefont
  {S.}~\bibnamefont {de~Gironcoli}}, \bibinfo {author} {\bibfnamefont
  {P.}~\bibnamefont {Delugas}}, \bibinfo {author} {\bibfnamefont {R.~A.}\
  \bibnamefont {DiStasio}}, \bibinfo {author} {\bibfnamefont {A.}~\bibnamefont
  {Ferretti}}, \bibinfo {author} {\bibfnamefont {A.}~\bibnamefont {Floris}},
  \bibinfo {author} {\bibfnamefont {G.}~\bibnamefont {Fratesi}}, \bibinfo
  {author} {\bibfnamefont {G.}~\bibnamefont {Fugallo}}, \bibinfo {author}
  {\bibfnamefont {R.}~\bibnamefont {Gebauer}}, \bibinfo {author} {\bibfnamefont
  {U.}~\bibnamefont {Gerstmann}}, \bibinfo {author} {\bibfnamefont
  {F.}~\bibnamefont {Giustino}}, \bibinfo {author} {\bibfnamefont
  {T.}~\bibnamefont {Gorni}}, \bibinfo {author} {\bibfnamefont
  {J.}~\bibnamefont {Jia}}, \bibinfo {author} {\bibfnamefont {M.}~\bibnamefont
  {Kawamura}}, \bibinfo {author} {\bibfnamefont {H.-Y.}\ \bibnamefont {Ko}},
  \bibinfo {author} {\bibfnamefont {A.}~\bibnamefont {Kokalj}}, \bibinfo
  {author} {\bibfnamefont {E.}~\bibnamefont {Küçükbenli}}, \bibinfo {author}
  {\bibfnamefont {M.}~\bibnamefont {Lazzeri}}, \bibinfo {author} {\bibfnamefont
  {M.}~\bibnamefont {Marsili}}, \bibinfo {author} {\bibfnamefont
  {N.}~\bibnamefont {Marzari}}, \bibinfo {author} {\bibfnamefont
  {F.}~\bibnamefont {Mauri}}, \bibinfo {author} {\bibfnamefont {N.~L.}\
  \bibnamefont {Nguyen}}, \bibinfo {author} {\bibfnamefont {H.-V.}\
  \bibnamefont {Nguyen}}, \bibinfo {author} {\bibfnamefont {A.}~\bibnamefont
  {Otero-de-la Roza}}, \bibinfo {author} {\bibfnamefont {L.}~\bibnamefont
  {Paulatto}}, \bibinfo {author} {\bibfnamefont {S.}~\bibnamefont {Poncé}},
  \bibinfo {author} {\bibfnamefont {D.}~\bibnamefont {Rocca}}, \bibinfo
  {author} {\bibfnamefont {R.}~\bibnamefont {Sabatini}}, \bibinfo {author}
  {\bibfnamefont {B.}~\bibnamefont {Santra}}, \bibinfo {author} {\bibfnamefont
  {M.}~\bibnamefont {Schlipf}}, \bibinfo {author} {\bibfnamefont {A.~P.}\
  \bibnamefont {Seitsonen}}, \bibinfo {author} {\bibfnamefont {A.}~\bibnamefont
  {Smogunov}}, \bibinfo {author} {\bibfnamefont {I.}~\bibnamefont {Timrov}},
  \bibinfo {author} {\bibfnamefont {T.}~\bibnamefont {Thonhauser}}, \bibinfo
  {author} {\bibfnamefont {P.}~\bibnamefont {Umari}}, \bibinfo {author}
  {\bibfnamefont {N.}~\bibnamefont {Vast}}, \bibinfo {author} {\bibfnamefont
  {X.}~\bibnamefont {Wu}}, \ and\ \bibinfo {author} {\bibfnamefont
  {S.}~\bibnamefont {Baroni}},\ }\href {\doibase 10.1088/1361-648X/aa8f79}
  {\bibfield  {journal} {\bibinfo  {journal} {Journal of Physics: Condensed
  Matter}\ }\textbf {\bibinfo {volume} {29}},\ \bibinfo {pages} {465901}
  (\bibinfo {year} {2017})}\BibitemShut {NoStop}%
\bibitem [{\citenamefont {Liechtenstein}\ \emph {et~al.}(1995)\citenamefont
  {Liechtenstein}, \citenamefont {Anisimov},\ and\ \citenamefont
  {Zaanen}}]{liechtenstein_density-functional_1995}%
  \BibitemOpen
  \bibfield  {author} {\bibinfo {author} {\bibfnamefont {A.~I.}\ \bibnamefont
  {Liechtenstein}}, \bibinfo {author} {\bibfnamefont {V.~I.}\ \bibnamefont
  {Anisimov}}, \ and\ \bibinfo {author} {\bibfnamefont {J.}~\bibnamefont
  {Zaanen}},\ }\href {\doibase 10.1103/PhysRevB.52.R5467} {\bibfield  {journal}
  {\bibinfo  {journal} {Physical Review B}\ }\textbf {\bibinfo {volume} {52}},\
  \bibinfo {pages} {R5467} (\bibinfo {year} {1995})}\BibitemShut {NoStop}%
\bibitem [{\citenamefont {Baroni}\ \emph {et~al.}(2001)\citenamefont {Baroni},
  \citenamefont {de~Gironcoli}, \citenamefont {Dal~Corso},\ and\ \citenamefont
  {Giannozzi}}]{baroni_phonons_2001}%
  \BibitemOpen
  \bibfield  {author} {\bibinfo {author} {\bibfnamefont {S.}~\bibnamefont
  {Baroni}}, \bibinfo {author} {\bibfnamefont {S.}~\bibnamefont
  {de~Gironcoli}}, \bibinfo {author} {\bibfnamefont {A.}~\bibnamefont
  {Dal~Corso}}, \ and\ \bibinfo {author} {\bibfnamefont {P.}~\bibnamefont
  {Giannozzi}},\ }\href {\doibase 10.1103/RevModPhys.73.515} {\bibfield
  {journal} {\bibinfo  {journal} {Reviews of Modern Physics}\ }\textbf
  {\bibinfo {volume} {73}},\ \bibinfo {pages} {515} (\bibinfo {year}
  {2001})}\BibitemShut {NoStop}%
\bibitem [{\citenamefont {Urru}\ and\ \citenamefont
  {Dal~Corso}(2019)}]{urru_density_2019}%
  \BibitemOpen
  \bibfield  {author} {\bibinfo {author} {\bibfnamefont {A.}~\bibnamefont
  {Urru}}\ and\ \bibinfo {author} {\bibfnamefont {A.}~\bibnamefont
  {Dal~Corso}},\ }\href {\doibase 10.1103/PhysRevB.100.045115} {\bibfield
  {journal} {\bibinfo  {journal} {Physical Review B}\ }\textbf {\bibinfo
  {volume} {100}},\ \bibinfo {pages} {045115} (\bibinfo {year}
  {2019})}\BibitemShut {NoStop}%
\bibitem [{\citenamefont {Dudarev}\ \emph {et~al.}(1998)\citenamefont
  {Dudarev}, \citenamefont {Botton}, \citenamefont {Savrasov}, \citenamefont
  {Humphreys},\ and\ \citenamefont
  {Sutton}}]{dudarev_electron-energy-loss_1998}%
  \BibitemOpen
  \bibfield  {author} {\bibinfo {author} {\bibfnamefont {S.~L.}\ \bibnamefont
  {Dudarev}}, \bibinfo {author} {\bibfnamefont {G.~A.}\ \bibnamefont {Botton}},
  \bibinfo {author} {\bibfnamefont {S.~Y.}\ \bibnamefont {Savrasov}}, \bibinfo
  {author} {\bibfnamefont {C.~J.}\ \bibnamefont {Humphreys}}, \ and\ \bibinfo
  {author} {\bibfnamefont {A.~P.}\ \bibnamefont {Sutton}},\ }\href {\doibase
  10.1103/PhysRevB.57.1505} {\bibfield  {journal} {\bibinfo  {journal}
  {Physical Review B}\ }\textbf {\bibinfo {volume} {57}},\ \bibinfo {pages}
  {1505} (\bibinfo {year} {1998})}\BibitemShut {NoStop}%
\bibitem [{\citenamefont {Timrov}\ \emph {et~al.}(2021)\citenamefont {Timrov},
  \citenamefont {Marzari},\ and\ \citenamefont
  {Cococcioni}}]{timrov_self-consistent_2021}%
  \BibitemOpen
  \bibfield  {author} {\bibinfo {author} {\bibfnamefont {I.}~\bibnamefont
  {Timrov}}, \bibinfo {author} {\bibfnamefont {N.}~\bibnamefont {Marzari}}, \
  and\ \bibinfo {author} {\bibfnamefont {M.}~\bibnamefont {Cococcioni}},\
  }\href {\doibase 10.1103/PhysRevB.103.045141} {\bibfield  {journal} {\bibinfo
   {journal} {Physical Review B}\ }\textbf {\bibinfo {volume} {103}},\ \bibinfo
  {pages} {045141} (\bibinfo {year} {2021})}\BibitemShut {NoStop}%
\bibitem [{\citenamefont {Cococcioni}(2010)}]{cococcioni_accurate_2010}%
  \BibitemOpen
  \bibfield  {author} {\bibinfo {author} {\bibfnamefont {M.}~\bibnamefont
  {Cococcioni}},\ }\href {\doibase 10.2138/rmg.2010.71.8} {\bibfield  {journal}
  {\bibinfo  {journal} {Reviews in Mineralogy and Geochemistry}\ }\textbf
  {\bibinfo {volume} {71}},\ \bibinfo {pages} {147} (\bibinfo {year}
  {2010})}\BibitemShut {NoStop}%
\bibitem [{\citenamefont
  {Dal~Corso}(2001)}]{dal_corso_density-functional_2001}%
  \BibitemOpen
  \bibfield  {author} {\bibinfo {author} {\bibfnamefont {A.}~\bibnamefont
  {Dal~Corso}},\ }\href {\doibase 10.1103/PhysRevB.64.235118} {\bibfield
  {journal} {\bibinfo  {journal} {Physical Review B}\ }\textbf {\bibinfo
  {volume} {64}},\ \bibinfo {pages} {235118} (\bibinfo {year}
  {2001})}\BibitemShut {NoStop}%
\bibitem [{\citenamefont {Timrov}\ \emph {et~al.}(2020)\citenamefont {Timrov},
  \citenamefont {Aquilante}, \citenamefont {Binci}, \citenamefont
  {Cococcioni},\ and\ \citenamefont {Marzari}}]{timrov_pulay_2020}%
  \BibitemOpen
  \bibfield  {author} {\bibinfo {author} {\bibfnamefont {I.}~\bibnamefont
  {Timrov}}, \bibinfo {author} {\bibfnamefont {F.}~\bibnamefont {Aquilante}},
  \bibinfo {author} {\bibfnamefont {L.}~\bibnamefont {Binci}}, \bibinfo
  {author} {\bibfnamefont {M.}~\bibnamefont {Cococcioni}}, \ and\ \bibinfo
  {author} {\bibfnamefont {N.}~\bibnamefont {Marzari}},\ }\href {\doibase
  10.1103/PhysRevB.102.235159} {\bibfield  {journal} {\bibinfo  {journal}
  {Physical Review B}\ }\textbf {\bibinfo {volume} {102}},\ \bibinfo {pages}
  {235159} (\bibinfo {year} {2020})}\BibitemShut {NoStop}%
\bibitem [{\citenamefont {Timrov}\ \emph {et~al.}(2018)\citenamefont {Timrov},
  \citenamefont {Marzari},\ and\ \citenamefont
  {Cococcioni}}]{timrov_hubbard_2018}%
  \BibitemOpen
  \bibfield  {author} {\bibinfo {author} {\bibfnamefont {I.}~\bibnamefont
  {Timrov}}, \bibinfo {author} {\bibfnamefont {N.}~\bibnamefont {Marzari}}, \
  and\ \bibinfo {author} {\bibfnamefont {M.}~\bibnamefont {Cococcioni}},\
  }\href {\doibase 10.1103/PhysRevB.98.085127} {\bibfield  {journal} {\bibinfo
  {journal} {Physical Review B}\ }\textbf {\bibinfo {volume} {98}},\ \bibinfo
  {pages} {085127} (\bibinfo {year} {2018})}\BibitemShut {NoStop}%
\bibitem [{\citenamefont {Cao}\ \emph {et~al.}(2018)\citenamefont {Cao},
  \citenamefont {Lambert}, \citenamefont {Radaelli},\ and\ \citenamefont
  {Giustino}}]{cao_ab_2018}%
  \BibitemOpen
  \bibfield  {author} {\bibinfo {author} {\bibfnamefont {K.}~\bibnamefont
  {Cao}}, \bibinfo {author} {\bibfnamefont {H.}~\bibnamefont {Lambert}},
  \bibinfo {author} {\bibfnamefont {P.~G.}\ \bibnamefont {Radaelli}}, \ and\
  \bibinfo {author} {\bibfnamefont {F.}~\bibnamefont {Giustino}},\ }\href
  {\doibase 10.1103/PhysRevB.97.024420} {\bibfield  {journal} {\bibinfo
  {journal} {Physical Review B}\ }\textbf {\bibinfo {volume} {97}},\ \bibinfo
  {pages} {024420} (\bibinfo {year} {2018})}\BibitemShut {NoStop}%
\bibitem [{\citenamefont {Gorni}\ \emph {et~al.}(2018)\citenamefont {Gorni},
  \citenamefont {Timrov},\ and\ \citenamefont {Baroni}}]{gorni_spin_2018}%
  \BibitemOpen
  \bibfield  {author} {\bibinfo {author} {\bibfnamefont {T.}~\bibnamefont
  {Gorni}}, \bibinfo {author} {\bibfnamefont {I.}~\bibnamefont {Timrov}}, \
  and\ \bibinfo {author} {\bibfnamefont {S.}~\bibnamefont {Baroni}},\ }\href
  {\doibase 10.1140/epjb/e2018-90247-9} {\bibfield  {journal} {\bibinfo
  {journal} {The European Physical Journal B}\ }\textbf {\bibinfo {volume}
  {91}},\ \bibinfo {pages} {249} (\bibinfo {year} {2018})}\BibitemShut
  {NoStop}%
\bibitem [{\citenamefont {Dal~Corso}(2007)}]{dal_corso_density_2007}%
  \BibitemOpen
  \bibfield  {author} {\bibinfo {author} {\bibfnamefont {A.}~\bibnamefont
  {Dal~Corso}},\ }\href {\doibase 10.1103/PhysRevB.76.054308} {\bibfield
  {journal} {\bibinfo  {journal} {Physical Review B}\ }\textbf {\bibinfo
  {volume} {76}},\ \bibinfo {pages} {054308} (\bibinfo {year}
  {2007})}\BibitemShut {NoStop}%
\bibitem [{\citenamefont {Dederichs}\ \emph {et~al.}(1984)\citenamefont
  {Dederichs}, \citenamefont {Blügel}, \citenamefont {Zeller},\ and\
  \citenamefont {Akai}}]{dederichs_ground_1984}%
  \BibitemOpen
  \bibfield  {author} {\bibinfo {author} {\bibfnamefont {P.~H.}\ \bibnamefont
  {Dederichs}}, \bibinfo {author} {\bibfnamefont {S.}~\bibnamefont {Blügel}},
  \bibinfo {author} {\bibfnamefont {R.}~\bibnamefont {Zeller}}, \ and\ \bibinfo
  {author} {\bibfnamefont {H.}~\bibnamefont {Akai}},\ }\href {\doibase
  10.1103/PhysRevLett.53.2512} {\bibfield  {journal} {\bibinfo  {journal}
  {Physical Review Letters}\ }\textbf {\bibinfo {volume} {53}},\ \bibinfo
  {pages} {2512} (\bibinfo {year} {1984})}\BibitemShut {NoStop}%
\bibitem [{\citenamefont {Hybertsen}\ \emph {et~al.}(1989)\citenamefont
  {Hybertsen}, \citenamefont {Schlüter},\ and\ \citenamefont
  {Christensen}}]{hybertsen_calculation_1989}%
  \BibitemOpen
  \bibfield  {author} {\bibinfo {author} {\bibfnamefont {M.~S.}\ \bibnamefont
  {Hybertsen}}, \bibinfo {author} {\bibfnamefont {M.}~\bibnamefont
  {Schlüter}}, \ and\ \bibinfo {author} {\bibfnamefont {N.~E.}\ \bibnamefont
  {Christensen}},\ }\href {\doibase 10.1103/PhysRevB.39.9028} {\bibfield
  {journal} {\bibinfo  {journal} {Physical Review B}\ }\textbf {\bibinfo
  {volume} {39}},\ \bibinfo {pages} {9028} (\bibinfo {year}
  {1989})}\BibitemShut {NoStop}%
\bibitem [{\citenamefont {Giuliani}\ and\ \citenamefont
  {Vignale}(2005)}]{giuliani_quantum_2005}%
  \BibitemOpen
  \bibfield  {author} {\bibinfo {author} {\bibfnamefont {G.}~\bibnamefont
  {Giuliani}}\ and\ \bibinfo {author} {\bibfnamefont {G.}~\bibnamefont
  {Vignale}},\ }\href {https://doi.org/10.1017/cbo9780511619915} {\emph
  {\bibinfo {title} {Quantum {Theory} of the {Electron} {Liquid}}}}\ (\bibinfo
  {publisher} {Cambridge University Press},\ \bibinfo {year}
  {2005})\BibitemShut {NoStop}%
\bibitem [{\citenamefont {Baroni}\ \emph {et~al.}(1987)\citenamefont {Baroni},
  \citenamefont {Giannozzi},\ and\ \citenamefont
  {Testa}}]{baroni_greens-function_1987}%
  \BibitemOpen
  \bibfield  {author} {\bibinfo {author} {\bibfnamefont {S.}~\bibnamefont
  {Baroni}}, \bibinfo {author} {\bibfnamefont {P.}~\bibnamefont {Giannozzi}}, \
  and\ \bibinfo {author} {\bibfnamefont {A.}~\bibnamefont {Testa}},\ }\href
  {\doibase 10.1103/PhysRevLett.58.1861} {\bibfield  {journal} {\bibinfo
  {journal} {Physical Review Letters}\ }\textbf {\bibinfo {volume} {58}},\
  \bibinfo {pages} {1861} (\bibinfo {year} {1987})}\BibitemShut {NoStop}%
\bibitem [{\citenamefont {de~Gironcoli}(1995)}]{de_gironcoli_lattice_1995}%
  \BibitemOpen
  \bibfield  {author} {\bibinfo {author} {\bibfnamefont {S.}~\bibnamefont
  {de~Gironcoli}},\ }\href {\doibase 10.1103/PhysRevB.51.6773} {\bibfield
  {journal} {\bibinfo  {journal} {Physical Review B}\ }\textbf {\bibinfo
  {volume} {51}},\ \bibinfo {pages} {6773} (\bibinfo {year}
  {1995})}\BibitemShut {NoStop}%
\bibitem [{\citenamefont {Giannozzi}\ \emph {et~al.}(2009)\citenamefont
  {Giannozzi}, \citenamefont {Baroni}, \citenamefont {Bonini}, \citenamefont
  {Calandra}, \citenamefont {Car}, \citenamefont {Cavazzoni}, \citenamefont
  {Ceresoli}, \citenamefont {Chiarotti}, \citenamefont {Cococcioni},
  \citenamefont {Dabo}, \citenamefont {Dal~Corso}, \citenamefont
  {de~Gironcoli}, \citenamefont {Fabris}, \citenamefont {Fratesi},
  \citenamefont {Gebauer}, \citenamefont {Gerstmann}, \citenamefont
  {Gougoussis}, \citenamefont {Kokalj}, \citenamefont {Lazzeri}, \citenamefont
  {Martin-Samos}, \citenamefont {Marzari}, \citenamefont {Mauri}, \citenamefont
  {Mazzarello}, \citenamefont {Paolini}, \citenamefont {Pasquarello},
  \citenamefont {Paulatto}, \citenamefont {Sbraccia}, \citenamefont {Scandolo},
  \citenamefont {Sclauzero}, \citenamefont {Seitsonen}, \citenamefont
  {Smogunov}, \citenamefont {Umari},\ and\ \citenamefont
  {Wentzcovitch}}]{giannozzi_quantum_2009}%
  \BibitemOpen
  \bibfield  {author} {\bibinfo {author} {\bibfnamefont {P.}~\bibnamefont
  {Giannozzi}}, \bibinfo {author} {\bibfnamefont {S.}~\bibnamefont {Baroni}},
  \bibinfo {author} {\bibfnamefont {N.}~\bibnamefont {Bonini}}, \bibinfo
  {author} {\bibfnamefont {M.}~\bibnamefont {Calandra}}, \bibinfo {author}
  {\bibfnamefont {R.}~\bibnamefont {Car}}, \bibinfo {author} {\bibfnamefont
  {C.}~\bibnamefont {Cavazzoni}}, \bibinfo {author} {\bibfnamefont
  {D.}~\bibnamefont {Ceresoli}}, \bibinfo {author} {\bibfnamefont {G.~L.}\
  \bibnamefont {Chiarotti}}, \bibinfo {author} {\bibfnamefont {M.}~\bibnamefont
  {Cococcioni}}, \bibinfo {author} {\bibfnamefont {I.}~\bibnamefont {Dabo}},
  \bibinfo {author} {\bibfnamefont {A.}~\bibnamefont {Dal~Corso}}, \bibinfo
  {author} {\bibfnamefont {S.}~\bibnamefont {de~Gironcoli}}, \bibinfo {author}
  {\bibfnamefont {S.}~\bibnamefont {Fabris}}, \bibinfo {author} {\bibfnamefont
  {G.}~\bibnamefont {Fratesi}}, \bibinfo {author} {\bibfnamefont
  {R.}~\bibnamefont {Gebauer}}, \bibinfo {author} {\bibfnamefont
  {U.}~\bibnamefont {Gerstmann}}, \bibinfo {author} {\bibfnamefont
  {C.}~\bibnamefont {Gougoussis}}, \bibinfo {author} {\bibfnamefont
  {A.}~\bibnamefont {Kokalj}}, \bibinfo {author} {\bibfnamefont
  {M.}~\bibnamefont {Lazzeri}}, \bibinfo {author} {\bibfnamefont
  {L.}~\bibnamefont {Martin-Samos}}, \bibinfo {author} {\bibfnamefont
  {N.}~\bibnamefont {Marzari}}, \bibinfo {author} {\bibfnamefont
  {F.}~\bibnamefont {Mauri}}, \bibinfo {author} {\bibfnamefont
  {R.}~\bibnamefont {Mazzarello}}, \bibinfo {author} {\bibfnamefont
  {S.}~\bibnamefont {Paolini}}, \bibinfo {author} {\bibfnamefont
  {A.}~\bibnamefont {Pasquarello}}, \bibinfo {author} {\bibfnamefont
  {L.}~\bibnamefont {Paulatto}}, \bibinfo {author} {\bibfnamefont
  {C.}~\bibnamefont {Sbraccia}}, \bibinfo {author} {\bibfnamefont
  {S.}~\bibnamefont {Scandolo}}, \bibinfo {author} {\bibfnamefont
  {G.}~\bibnamefont {Sclauzero}}, \bibinfo {author} {\bibfnamefont {A.~P.}\
  \bibnamefont {Seitsonen}}, \bibinfo {author} {\bibfnamefont {A.}~\bibnamefont
  {Smogunov}}, \bibinfo {author} {\bibfnamefont {P.}~\bibnamefont {Umari}}, \
  and\ \bibinfo {author} {\bibfnamefont {R.~M.}\ \bibnamefont {Wentzcovitch}},\
  }\href {\doibase 10.1088/0953-8984/21/39/395502} {\bibfield  {journal}
  {\bibinfo  {journal} {Journal of Physics: Condensed Matter}\ }\textbf
  {\bibinfo {volume} {21}},\ \bibinfo {pages} {395502} (\bibinfo {year}
  {2009})}\BibitemShut {NoStop}%
\bibitem [{\citenamefont {Timrov}\ \emph {et~al.}(2022)\citenamefont {Timrov},
  \citenamefont {Marzari},\ and\ \citenamefont {Cococcioni}}]{timrov_hp_2022}%
  \BibitemOpen
  \bibfield  {author} {\bibinfo {author} {\bibfnamefont {I.}~\bibnamefont
  {Timrov}}, \bibinfo {author} {\bibfnamefont {N.}~\bibnamefont {Marzari}}, \
  and\ \bibinfo {author} {\bibfnamefont {M.}~\bibnamefont {Cococcioni}},\
  }\href {\doibase 10.1016/j.cpc.2022.108455} {\bibfield  {journal} {\bibinfo
  {journal} {Computer Physics Communications}\ }\textbf {\bibinfo {volume}
  {279}},\ \bibinfo {pages} {108455} (\bibinfo {year} {2022})}\BibitemShut
  {NoStop}%
\bibitem [{\citenamefont {Dal~Corso}(2014)}]{dal_corso_pseudopotentials_2014}%
  \BibitemOpen
  \bibfield  {author} {\bibinfo {author} {\bibfnamefont {A.}~\bibnamefont
  {Dal~Corso}},\ }\href {\doibase 10.1016/j.commatsci.2014.07.043} {\bibfield
  {journal} {\bibinfo  {journal} {Computational Materials Science}\ }\textbf
  {\bibinfo {volume} {95}},\ \bibinfo {pages} {337} (\bibinfo {year}
  {2014})}\BibitemShut {NoStop}%
\bibitem [{\citenamefont {Perdew}\ \emph {et~al.}(2008)\citenamefont {Perdew},
  \citenamefont {Ruzsinszky}, \citenamefont {Csonka}, \citenamefont {Vydrov},
  \citenamefont {Scuseria}, \citenamefont {Constantin}, \citenamefont {Zhou},\
  and\ \citenamefont {Burke}}]{perdew_restoring_2008}%
  \BibitemOpen
  \bibfield  {author} {\bibinfo {author} {\bibfnamefont {J.~P.}\ \bibnamefont
  {Perdew}}, \bibinfo {author} {\bibfnamefont {A.}~\bibnamefont {Ruzsinszky}},
  \bibinfo {author} {\bibfnamefont {G.~I.}\ \bibnamefont {Csonka}}, \bibinfo
  {author} {\bibfnamefont {O.~A.}\ \bibnamefont {Vydrov}}, \bibinfo {author}
  {\bibfnamefont {G.~E.}\ \bibnamefont {Scuseria}}, \bibinfo {author}
  {\bibfnamefont {L.~A.}\ \bibnamefont {Constantin}}, \bibinfo {author}
  {\bibfnamefont {X.}~\bibnamefont {Zhou}}, \ and\ \bibinfo {author}
  {\bibfnamefont {K.}~\bibnamefont {Burke}},\ }\href {\doibase
  10.1103/PhysRevLett.100.136406} {\bibfield  {journal} {\bibinfo  {journal}
  {Physical Review Letters}\ }\textbf {\bibinfo {volume} {100}},\ \bibinfo
  {pages} {136406} (\bibinfo {year} {2008})}\BibitemShut {NoStop}%
\bibitem [{\citenamefont {Jain}\ \emph {et~al.}(2013)\citenamefont {Jain},
  \citenamefont {Ong}, \citenamefont {Hautier}, \citenamefont {Chen},
  \citenamefont {Richards}, \citenamefont {Dacek}, \citenamefont {Cholia},
  \citenamefont {Gunter}, \citenamefont {Skinner}, \citenamefont {Ceder},\ and\
  \citenamefont {Persson}}]{jain_commentary_2013}%
  \BibitemOpen
  \bibfield  {author} {\bibinfo {author} {\bibfnamefont {A.}~\bibnamefont
  {Jain}}, \bibinfo {author} {\bibfnamefont {S.~P.}\ \bibnamefont {Ong}},
  \bibinfo {author} {\bibfnamefont {G.}~\bibnamefont {Hautier}}, \bibinfo
  {author} {\bibfnamefont {W.}~\bibnamefont {Chen}}, \bibinfo {author}
  {\bibfnamefont {W.~D.}\ \bibnamefont {Richards}}, \bibinfo {author}
  {\bibfnamefont {S.}~\bibnamefont {Dacek}}, \bibinfo {author} {\bibfnamefont
  {S.}~\bibnamefont {Cholia}}, \bibinfo {author} {\bibfnamefont
  {D.}~\bibnamefont {Gunter}}, \bibinfo {author} {\bibfnamefont
  {D.}~\bibnamefont {Skinner}}, \bibinfo {author} {\bibfnamefont
  {G.}~\bibnamefont {Ceder}}, \ and\ \bibinfo {author} {\bibfnamefont {K.~A.}\
  \bibnamefont {Persson}},\ }\href {\doibase 10.1063/1.4812323} {\bibfield
  {journal} {\bibinfo  {journal} {APL Materials}\ }\textbf {\bibinfo {volume}
  {1}},\ \bibinfo {pages} {011002} (\bibinfo {year} {2013})}\BibitemShut
  {NoStop}%
\bibitem [{\citenamefont {Hinuma}\ \emph {et~al.}(2017)\citenamefont {Hinuma},
  \citenamefont {Pizzi}, \citenamefont {Kumagai}, \citenamefont {Oba},\ and\
  \citenamefont {Tanaka}}]{hinuma_band_2017}%
  \BibitemOpen
  \bibfield  {author} {\bibinfo {author} {\bibfnamefont {Y.}~\bibnamefont
  {Hinuma}}, \bibinfo {author} {\bibfnamefont {G.}~\bibnamefont {Pizzi}},
  \bibinfo {author} {\bibfnamefont {Y.}~\bibnamefont {Kumagai}}, \bibinfo
  {author} {\bibfnamefont {F.}~\bibnamefont {Oba}}, \ and\ \bibinfo {author}
  {\bibfnamefont {I.}~\bibnamefont {Tanaka}},\ }\href {\doibase
  10.1016/j.commatsci.2016.10.015} {\bibfield  {journal} {\bibinfo  {journal}
  {Computational Materials Science}\ }\textbf {\bibinfo {volume} {128}},\
  \bibinfo {pages} {140} (\bibinfo {year} {2017})}\BibitemShut {NoStop}%
\bibitem [{\citenamefont {Campbell}\ \emph {et~al.}(2006)\citenamefont
  {Campbell}, \citenamefont {Stokes}, \citenamefont {Tanner},\ and\
  \citenamefont {Hatch}}]{campbell_isodisplace_2006}%
  \BibitemOpen
  \bibfield  {author} {\bibinfo {author} {\bibfnamefont {B.~J.}\ \bibnamefont
  {Campbell}}, \bibinfo {author} {\bibfnamefont {H.~T.}\ \bibnamefont
  {Stokes}}, \bibinfo {author} {\bibfnamefont {D.~E.}\ \bibnamefont {Tanner}},
  \ and\ \bibinfo {author} {\bibfnamefont {D.~M.}\ \bibnamefont {Hatch}},\
  }\href {\doibase 10.1107/S0021889806014075} {\bibfield  {journal} {\bibinfo
  {journal} {Journal of Applied Crystallography}\ }\textbf {\bibinfo {volume}
  {39}},\ \bibinfo {pages} {607} (\bibinfo {year} {2006})}\BibitemShut
  {NoStop}%
\bibitem [{\citenamefont {Momma}\ and\ \citenamefont
  {Izumi}(2008)}]{momma_vesta_2008}%
  \BibitemOpen
  \bibfield  {author} {\bibinfo {author} {\bibfnamefont {K.}~\bibnamefont
  {Momma}}\ and\ \bibinfo {author} {\bibfnamefont {F.}~\bibnamefont {Izumi}},\
  }\href {\doibase 10.1107/S0021889808012016} {\bibfield  {journal} {\bibinfo
  {journal} {Journal of Applied Crystallography}\ }\textbf {\bibinfo {volume}
  {41}},\ \bibinfo {pages} {653} (\bibinfo {year} {2008})}\BibitemShut
  {NoStop}%
\bibitem [{\citenamefont {Binci}\ and\ \citenamefont
  {Marzari}()}]{MaterialsCloudArchive2023}%
  \BibitemOpen
  \bibfield  {author} {\bibinfo {author} {\bibfnamefont {L.}~\bibnamefont
  {Binci}}\ and\ \bibinfo {author} {\bibfnamefont {N.}~\bibnamefont
  {Marzari}},\ }\href {\doibase 10.24435/materialscloud:3d-ww} {}\bibinfo
  {howpublished} {Materials Cloud Archive {\bf 2023.67} (2023), doi:
  10.24435/materialscloud:3d-ww}\BibitemShut {NoStop}%
\bibitem [{\citenamefont {Monteiro}\ \emph {et~al.}(2013)\citenamefont
  {Monteiro}, \citenamefont {Baker}, \citenamefont {Ionescu}, \citenamefont
  {Barnes}, \citenamefont {Salman}, \citenamefont {Suter}, \citenamefont
  {Prokscha},\ and\ \citenamefont {Langridge}}]{monteiro_spatially_2013}%
  \BibitemOpen
  \bibfield  {author} {\bibinfo {author} {\bibfnamefont {P.~M.~S.}\
  \bibnamefont {Monteiro}}, \bibinfo {author} {\bibfnamefont {P.~J.}\
  \bibnamefont {Baker}}, \bibinfo {author} {\bibfnamefont {A.}~\bibnamefont
  {Ionescu}}, \bibinfo {author} {\bibfnamefont {C.~H.~W.}\ \bibnamefont
  {Barnes}}, \bibinfo {author} {\bibfnamefont {Z.}~\bibnamefont {Salman}},
  \bibinfo {author} {\bibfnamefont {A.}~\bibnamefont {Suter}}, \bibinfo
  {author} {\bibfnamefont {T.}~\bibnamefont {Prokscha}}, \ and\ \bibinfo
  {author} {\bibfnamefont {S.}~\bibnamefont {Langridge}},\ }\href {\doibase
  10.1103/PhysRevLett.110.217208} {\bibfield  {journal} {\bibinfo  {journal}
  {Physical Review Letters}\ }\textbf {\bibinfo {volume} {110}},\ \bibinfo
  {pages} {217208} (\bibinfo {year} {2013})}\BibitemShut {NoStop}%
\bibitem [{\citenamefont {Mauger}\ and\ \citenamefont
  {Godart}(1986)}]{mauger_magnetic_1986}%
  \BibitemOpen
  \bibfield  {author} {\bibinfo {author} {\bibfnamefont {A.}~\bibnamefont
  {Mauger}}\ and\ \bibinfo {author} {\bibfnamefont {C.}~\bibnamefont
  {Godart}},\ }\href {\doibase 10.1016/0370-1573(86)90139-0} {\bibfield
  {journal} {\bibinfo  {journal} {Physics Reports}\ }\textbf {\bibinfo {volume}
  {141}},\ \bibinfo {pages} {51} (\bibinfo {year} {1986})}\BibitemShut
  {NoStop}%
\bibitem [{\citenamefont {Fujiwara}\ \emph {et~al.}(1982)\citenamefont
  {Fujiwara}, \citenamefont {Kadomatsu}, \citenamefont {Kurisu}, \citenamefont
  {Hihara}, \citenamefont {Kojima},\ and\ \citenamefont
  {Kamigaichi}}]{fujiwara_pressure-induced_1982}%
  \BibitemOpen
  \bibfield  {author} {\bibinfo {author} {\bibfnamefont {H.}~\bibnamefont
  {Fujiwara}}, \bibinfo {author} {\bibfnamefont {H.}~\bibnamefont {Kadomatsu}},
  \bibinfo {author} {\bibfnamefont {M.}~\bibnamefont {Kurisu}}, \bibinfo
  {author} {\bibfnamefont {T.}~\bibnamefont {Hihara}}, \bibinfo {author}
  {\bibfnamefont {K.}~\bibnamefont {Kojima}}, \ and\ \bibinfo {author}
  {\bibfnamefont {T.}~\bibnamefont {Kamigaichi}},\ }\href {\doibase
  10.1016/0038-1098(82)90631-7} {\bibfield  {journal} {\bibinfo  {journal}
  {Solid State Communications}\ }\textbf {\bibinfo {volume} {42}},\ \bibinfo
  {pages} {509} (\bibinfo {year} {1982})}\BibitemShut {NoStop}%
\bibitem [{\citenamefont {Ishizuka}\ \emph {et~al.}(1997)\citenamefont
  {Ishizuka}, \citenamefont {Kai}, \citenamefont {Akimoto}, \citenamefont
  {Kobayashi}, \citenamefont {Amaya},\ and\ \citenamefont
  {Endo}}]{ishizuka_pressure-induced_1997}%
  \BibitemOpen
  \bibfield  {author} {\bibinfo {author} {\bibfnamefont {M.}~\bibnamefont
  {Ishizuka}}, \bibinfo {author} {\bibfnamefont {Y.}~\bibnamefont {Kai}},
  \bibinfo {author} {\bibfnamefont {R.}~\bibnamefont {Akimoto}}, \bibinfo
  {author} {\bibfnamefont {M.}~\bibnamefont {Kobayashi}}, \bibinfo {author}
  {\bibfnamefont {K.}~\bibnamefont {Amaya}}, \ and\ \bibinfo {author}
  {\bibfnamefont {S.}~\bibnamefont {Endo}},\ }\href {\doibase
  10.1016/S0304-8853(96)00432-5} {\bibfield  {journal} {\bibinfo  {journal}
  {Journal of Magnetism and Magnetic Materials}\ }\textbf {\bibinfo {volume}
  {166}},\ \bibinfo {pages} {211} (\bibinfo {year} {1997})}\BibitemShut
  {NoStop}%
\bibitem [{\citenamefont {Jayaraman}\ \emph {et~al.}(1974)\citenamefont
  {Jayaraman}, \citenamefont {Singh}, \citenamefont {Chatterjee},\ and\
  \citenamefont {Devi}}]{jayaraman_pressure-volume_1974}%
  \BibitemOpen
  \bibfield  {author} {\bibinfo {author} {\bibfnamefont {A.}~\bibnamefont
  {Jayaraman}}, \bibinfo {author} {\bibfnamefont {A.~K.}\ \bibnamefont
  {Singh}}, \bibinfo {author} {\bibfnamefont {A.}~\bibnamefont {Chatterjee}}, \
  and\ \bibinfo {author} {\bibfnamefont {S.~U.}\ \bibnamefont {Devi}},\ }\href
  {\doibase 10.1103/PhysRevB.9.2513} {\bibfield  {journal} {\bibinfo  {journal}
  {Physical Review B}\ }\textbf {\bibinfo {volume} {9}},\ \bibinfo {pages}
  {2513} (\bibinfo {year} {1974})}\BibitemShut {NoStop}%
\bibitem [{\citenamefont {Barbagallo}\ \emph {et~al.}(2010)\citenamefont
  {Barbagallo}, \citenamefont {Hine}, \citenamefont {Cooper}, \citenamefont
  {Steinke}, \citenamefont {Ionescu}, \citenamefont {Barnes}, \citenamefont
  {Kinane}, \citenamefont {Dalgliesh}, \citenamefont {Charlton},\ and\
  \citenamefont {Langridge}}]{barbagallo_experimental_2010}%
  \BibitemOpen
  \bibfield  {author} {\bibinfo {author} {\bibfnamefont {M.}~\bibnamefont
  {Barbagallo}}, \bibinfo {author} {\bibfnamefont {N.~D.~M.}\ \bibnamefont
  {Hine}}, \bibinfo {author} {\bibfnamefont {J.~F.~K.}\ \bibnamefont {Cooper}},
  \bibinfo {author} {\bibfnamefont {N.-J.}\ \bibnamefont {Steinke}}, \bibinfo
  {author} {\bibfnamefont {A.}~\bibnamefont {Ionescu}}, \bibinfo {author}
  {\bibfnamefont {C.~H.~W.}\ \bibnamefont {Barnes}}, \bibinfo {author}
  {\bibfnamefont {C.~J.}\ \bibnamefont {Kinane}}, \bibinfo {author}
  {\bibfnamefont {R.~M.}\ \bibnamefont {Dalgliesh}}, \bibinfo {author}
  {\bibfnamefont {T.~R.}\ \bibnamefont {Charlton}}, \ and\ \bibinfo {author}
  {\bibfnamefont {S.}~\bibnamefont {Langridge}},\ }\href {\doibase
  10.1103/PhysRevB.81.235216} {\bibfield  {journal} {\bibinfo  {journal}
  {Physical Review B}\ }\textbf {\bibinfo {volume} {81}},\ \bibinfo {pages}
  {235216} (\bibinfo {year} {2010})}\BibitemShut {NoStop}%
\bibitem [{\citenamefont {Wan}\ \emph {et~al.}(2011{\natexlab{b}})\citenamefont
  {Wan}, \citenamefont {Dong},\ and\ \citenamefont
  {Savrasov}}]{wan_mechanism_2011}%
  \BibitemOpen
  \bibfield  {author} {\bibinfo {author} {\bibfnamefont {X.}~\bibnamefont
  {Wan}}, \bibinfo {author} {\bibfnamefont {J.}~\bibnamefont {Dong}}, \ and\
  \bibinfo {author} {\bibfnamefont {S.~Y.}\ \bibnamefont {Savrasov}},\ }\href
  {\doibase 10.1103/PhysRevB.83.205201} {\bibfield  {journal} {\bibinfo
  {journal} {Physical Review B}\ }\textbf {\bibinfo {volume} {83}},\ \bibinfo
  {pages} {205201} (\bibinfo {year} {2011}{\natexlab{b}})}\BibitemShut
  {NoStop}%
\bibitem [{\citenamefont {Kunes}\ \emph {et~al.}(2005)\citenamefont {Kunes},
  \citenamefont {Ku},\ and\ \citenamefont {E.~Pickett}}]{kunes_exchange_2005}%
  \BibitemOpen
  \bibfield  {author} {\bibinfo {author} {\bibfnamefont {J.}~\bibnamefont
  {Kunes}}, \bibinfo {author} {\bibfnamefont {W.}~\bibnamefont {Ku}}, \ and\
  \bibinfo {author} {\bibfnamefont {W.}~\bibnamefont {E.~Pickett}},\ }\href
  {\doibase 10.1143/JPSJ.74.1408} {\bibfield  {journal} {\bibinfo  {journal}
  {Journal of the Physical Society of Japan}\ }\textbf {\bibinfo {volume}
  {74}},\ \bibinfo {pages} {1408} (\bibinfo {year} {2005})}\BibitemShut
  {NoStop}%
\bibitem [{\citenamefont {Gardner}\ \emph {et~al.}(2010)\citenamefont
  {Gardner}, \citenamefont {Gingras},\ and\ \citenamefont
  {Greedan}}]{gardner_magnetic_2010}%
  \BibitemOpen
  \bibfield  {author} {\bibinfo {author} {\bibfnamefont {J.~S.}\ \bibnamefont
  {Gardner}}, \bibinfo {author} {\bibfnamefont {M.~J.~P.}\ \bibnamefont
  {Gingras}}, \ and\ \bibinfo {author} {\bibfnamefont {J.~E.}\ \bibnamefont
  {Greedan}},\ }\href {\doibase 10.1103/RevModPhys.82.53} {\bibfield  {journal}
  {\bibinfo  {journal} {Reviews of Modern Physics}\ }\textbf {\bibinfo {volume}
  {82}},\ \bibinfo {pages} {53} (\bibinfo {year} {2010})}\BibitemShut {NoStop}%
\bibitem [{\citenamefont {Harris}\ \emph {et~al.}(1997)\citenamefont {Harris},
  \citenamefont {Bramwell}, \citenamefont {McMorrow}, \citenamefont {Zeiske},\
  and\ \citenamefont {Godfrey}}]{harris_geometrical_1997}%
  \BibitemOpen
  \bibfield  {author} {\bibinfo {author} {\bibfnamefont {M.~J.}\ \bibnamefont
  {Harris}}, \bibinfo {author} {\bibfnamefont {S.~T.}\ \bibnamefont
  {Bramwell}}, \bibinfo {author} {\bibfnamefont {D.~F.}\ \bibnamefont
  {McMorrow}}, \bibinfo {author} {\bibfnamefont {T.}~\bibnamefont {Zeiske}}, \
  and\ \bibinfo {author} {\bibfnamefont {K.~W.}\ \bibnamefont {Godfrey}},\
  }\href {\doibase 10.1103/PhysRevLett.79.2554} {\bibfield  {journal} {\bibinfo
   {journal} {Physical Review Letters}\ }\textbf {\bibinfo {volume} {79}},\
  \bibinfo {pages} {2554} (\bibinfo {year} {1997})}\BibitemShut {NoStop}%
\bibitem [{\citenamefont {Ramirez}\ \emph {et~al.}(1999)\citenamefont
  {Ramirez}, \citenamefont {Hayashi}, \citenamefont {Cava}, \citenamefont
  {Siddharthan},\ and\ \citenamefont {Shastry}}]{ramirez_zero-point_1999}%
  \BibitemOpen
  \bibfield  {author} {\bibinfo {author} {\bibfnamefont {A.~P.}\ \bibnamefont
  {Ramirez}}, \bibinfo {author} {\bibfnamefont {A.}~\bibnamefont {Hayashi}},
  \bibinfo {author} {\bibfnamefont {R.~J.}\ \bibnamefont {Cava}}, \bibinfo
  {author} {\bibfnamefont {R.}~\bibnamefont {Siddharthan}}, \ and\ \bibinfo
  {author} {\bibfnamefont {B.~S.}\ \bibnamefont {Shastry}},\ }\href {\doibase
  10.1038/20619} {\bibfield  {journal} {\bibinfo  {journal} {Nature}\ }\textbf
  {\bibinfo {volume} {399}},\ \bibinfo {pages} {333} (\bibinfo {year}
  {1999})}\BibitemShut {NoStop}%
\bibitem [{\citenamefont {Calder}\ \emph {et~al.}(2016)\citenamefont {Calder},
  \citenamefont {Vale}, \citenamefont {Bogdanov}, \citenamefont {Liu},
  \citenamefont {Donnerer}, \citenamefont {Upton}, \citenamefont {Casa},
  \citenamefont {Said}, \citenamefont {Lumsden}, \citenamefont {Zhao},
  \citenamefont {Yan}, \citenamefont {Mandrus}, \citenamefont {Nishimoto},
  \citenamefont {van~den Brink}, \citenamefont {Hill}, \citenamefont
  {McMorrow},\ and\ \citenamefont
  {Christianson}}]{calder_spin-orbit-driven_2016}%
  \BibitemOpen
  \bibfield  {author} {\bibinfo {author} {\bibfnamefont {S.}~\bibnamefont
  {Calder}}, \bibinfo {author} {\bibfnamefont {J.~G.}\ \bibnamefont {Vale}},
  \bibinfo {author} {\bibfnamefont {N.~A.}\ \bibnamefont {Bogdanov}}, \bibinfo
  {author} {\bibfnamefont {X.}~\bibnamefont {Liu}}, \bibinfo {author}
  {\bibfnamefont {C.}~\bibnamefont {Donnerer}}, \bibinfo {author}
  {\bibfnamefont {M.~H.}\ \bibnamefont {Upton}}, \bibinfo {author}
  {\bibfnamefont {D.}~\bibnamefont {Casa}}, \bibinfo {author} {\bibfnamefont
  {A.~H.}\ \bibnamefont {Said}}, \bibinfo {author} {\bibfnamefont {M.~D.}\
  \bibnamefont {Lumsden}}, \bibinfo {author} {\bibfnamefont {Z.}~\bibnamefont
  {Zhao}}, \bibinfo {author} {\bibfnamefont {J.~Q.}\ \bibnamefont {Yan}},
  \bibinfo {author} {\bibfnamefont {D.}~\bibnamefont {Mandrus}}, \bibinfo
  {author} {\bibfnamefont {S.}~\bibnamefont {Nishimoto}}, \bibinfo {author}
  {\bibfnamefont {J.}~\bibnamefont {van~den Brink}}, \bibinfo {author}
  {\bibfnamefont {J.~P.}\ \bibnamefont {Hill}}, \bibinfo {author}
  {\bibfnamefont {D.~F.}\ \bibnamefont {McMorrow}}, \ and\ \bibinfo {author}
  {\bibfnamefont {A.~D.}\ \bibnamefont {Christianson}},\ }\href {\doibase
  10.1038/ncomms11651} {\bibfield  {journal} {\bibinfo  {journal} {Nature
  Communications}\ }\textbf {\bibinfo {volume} {7}},\ \bibinfo {pages} {11651}
  (\bibinfo {year} {2016})}\BibitemShut {NoStop}%
\bibitem [{\citenamefont {Sohn}\ \emph {et~al.}(2015)\citenamefont {Sohn},
  \citenamefont {Jeong}, \citenamefont {Jin}, \citenamefont {Kim},
  \citenamefont {Sandilands}, \citenamefont {Park}, \citenamefont {Kim},
  \citenamefont {Moon}, \citenamefont {Cho}, \citenamefont {Yamaura},
  \citenamefont {Hiroi},\ and\ \citenamefont {Noh}}]{sohn_optical_2015}%
  \BibitemOpen
  \bibfield  {author} {\bibinfo {author} {\bibfnamefont {C.}~\bibnamefont
  {Sohn}}, \bibinfo {author} {\bibfnamefont {H.}~\bibnamefont {Jeong}},
  \bibinfo {author} {\bibfnamefont {H.}~\bibnamefont {Jin}}, \bibinfo {author}
  {\bibfnamefont {S.}~\bibnamefont {Kim}}, \bibinfo {author} {\bibfnamefont
  {L.}~\bibnamefont {Sandilands}}, \bibinfo {author} {\bibfnamefont
  {H.}~\bibnamefont {Park}}, \bibinfo {author} {\bibfnamefont {K.}~\bibnamefont
  {Kim}}, \bibinfo {author} {\bibfnamefont {S.}~\bibnamefont {Moon}}, \bibinfo
  {author} {\bibfnamefont {D.-Y.}\ \bibnamefont {Cho}}, \bibinfo {author}
  {\bibfnamefont {J.}~\bibnamefont {Yamaura}}, \bibinfo {author} {\bibfnamefont
  {Z.}~\bibnamefont {Hiroi}}, \ and\ \bibinfo {author} {\bibfnamefont
  {T.}~\bibnamefont {Noh}},\ }\href {\doibase 10.1103/PhysRevLett.115.266402}
  {\bibfield  {journal} {\bibinfo  {journal} {Physical Review Letters}\
  }\textbf {\bibinfo {volume} {115}},\ \bibinfo {pages} {266402} (\bibinfo
  {year} {2015})}\BibitemShut {NoStop}%
\bibitem [{\citenamefont {Mandrus}\ \emph {et~al.}(2001)\citenamefont
  {Mandrus}, \citenamefont {Thompson}, \citenamefont {Gaal}, \citenamefont
  {Forro}, \citenamefont {Bryan}, \citenamefont {Chakoumakos}, \citenamefont
  {Woods}, \citenamefont {Sales}, \citenamefont {Fishman},\ and\ \citenamefont
  {Keppens}}]{mandrus_continuous_2001}%
  \BibitemOpen
  \bibfield  {author} {\bibinfo {author} {\bibfnamefont {D.}~\bibnamefont
  {Mandrus}}, \bibinfo {author} {\bibfnamefont {J.~R.}\ \bibnamefont
  {Thompson}}, \bibinfo {author} {\bibfnamefont {R.}~\bibnamefont {Gaal}},
  \bibinfo {author} {\bibfnamefont {L.}~\bibnamefont {Forro}}, \bibinfo
  {author} {\bibfnamefont {J.~C.}\ \bibnamefont {Bryan}}, \bibinfo {author}
  {\bibfnamefont {B.~C.}\ \bibnamefont {Chakoumakos}}, \bibinfo {author}
  {\bibfnamefont {L.~M.}\ \bibnamefont {Woods}}, \bibinfo {author}
  {\bibfnamefont {B.~C.}\ \bibnamefont {Sales}}, \bibinfo {author}
  {\bibfnamefont {R.~S.}\ \bibnamefont {Fishman}}, \ and\ \bibinfo {author}
  {\bibfnamefont {V.}~\bibnamefont {Keppens}},\ }\href {\doibase
  10.1103/PhysRevB.63.195104} {\bibfield  {journal} {\bibinfo  {journal}
  {Physical Review B}\ }\textbf {\bibinfo {volume} {63}},\ \bibinfo {pages}
  {195104} (\bibinfo {year} {2001})}\BibitemShut {NoStop}%
\bibitem [{\citenamefont {Yamaura}\ \emph {et~al.}(2012)\citenamefont
  {Yamaura}, \citenamefont {Ohgushi}, \citenamefont {Ohsumi}, \citenamefont
  {Hasegawa}, \citenamefont {Yamauchi}, \citenamefont {Sugimoto}, \citenamefont
  {Takeshita}, \citenamefont {Tokuda}, \citenamefont {Takata}, \citenamefont
  {Udagawa}, \citenamefont {Takigawa}, \citenamefont {Harima}, \citenamefont
  {Arima},\ and\ \citenamefont {Hiroi}}]{yamaura_tetrahedral_2012}%
  \BibitemOpen
  \bibfield  {author} {\bibinfo {author} {\bibfnamefont {J.}~\bibnamefont
  {Yamaura}}, \bibinfo {author} {\bibfnamefont {K.}~\bibnamefont {Ohgushi}},
  \bibinfo {author} {\bibfnamefont {H.}~\bibnamefont {Ohsumi}}, \bibinfo
  {author} {\bibfnamefont {T.}~\bibnamefont {Hasegawa}}, \bibinfo {author}
  {\bibfnamefont {I.}~\bibnamefont {Yamauchi}}, \bibinfo {author}
  {\bibfnamefont {K.}~\bibnamefont {Sugimoto}}, \bibinfo {author}
  {\bibfnamefont {S.}~\bibnamefont {Takeshita}}, \bibinfo {author}
  {\bibfnamefont {A.}~\bibnamefont {Tokuda}}, \bibinfo {author} {\bibfnamefont
  {M.}~\bibnamefont {Takata}}, \bibinfo {author} {\bibfnamefont
  {M.}~\bibnamefont {Udagawa}}, \bibinfo {author} {\bibfnamefont
  {M.}~\bibnamefont {Takigawa}}, \bibinfo {author} {\bibfnamefont
  {H.}~\bibnamefont {Harima}}, \bibinfo {author} {\bibfnamefont
  {T.}~\bibnamefont {Arima}}, \ and\ \bibinfo {author} {\bibfnamefont
  {Z.}~\bibnamefont {Hiroi}},\ }\href {\doibase 10.1103/PhysRevLett.108.247205}
  {\bibfield  {journal} {\bibinfo  {journal} {Physical Review Letters}\
  }\textbf {\bibinfo {volume} {108}},\ \bibinfo {pages} {247205} (\bibinfo
  {year} {2012})}\BibitemShut {NoStop}%
\bibitem [{\citenamefont {Kim}\ \emph {et~al.}(2020)\citenamefont {Kim},
  \citenamefont {Kim}, \citenamefont {Jeong}, \citenamefont {Park},
  \citenamefont {Park}, \citenamefont {Lee}, \citenamefont {Leiner},
  \citenamefont {Ishikawa}, \citenamefont {Baron}, \citenamefont {Hiroi},\ and\
  \citenamefont {Park}}]{kim_spin-orbit_2020}%
  \BibitemOpen
  \bibfield  {author} {\bibinfo {author} {\bibfnamefont {T.}~\bibnamefont
  {Kim}}, \bibinfo {author} {\bibfnamefont {C.~H.}\ \bibnamefont {Kim}},
  \bibinfo {author} {\bibfnamefont {J.}~\bibnamefont {Jeong}}, \bibinfo
  {author} {\bibfnamefont {P.}~\bibnamefont {Park}}, \bibinfo {author}
  {\bibfnamefont {K.}~\bibnamefont {Park}}, \bibinfo {author} {\bibfnamefont
  {K.~H.}\ \bibnamefont {Lee}}, \bibinfo {author} {\bibfnamefont {J.~C.}\
  \bibnamefont {Leiner}}, \bibinfo {author} {\bibfnamefont {D.}~\bibnamefont
  {Ishikawa}}, \bibinfo {author} {\bibfnamefont {A.~Q.~R.}\ \bibnamefont
  {Baron}}, \bibinfo {author} {\bibfnamefont {Z.}~\bibnamefont {Hiroi}}, \ and\
  \bibinfo {author} {\bibfnamefont {J.-G.}\ \bibnamefont {Park}},\ }\href
  {\doibase 10.1103/PhysRevB.102.201101} {\bibfield  {journal} {\bibinfo
  {journal} {Physical Review B}\ }\textbf {\bibinfo {volume} {102}},\ \bibinfo
  {pages} {201101} (\bibinfo {year} {2020})}\BibitemShut {NoStop}%
\bibitem [{\citenamefont {Shinaoka}\ \emph {et~al.}(2012)\citenamefont
  {Shinaoka}, \citenamefont {Miyake},\ and\ \citenamefont
  {Ishibashi}}]{shinaoka_noncollinear_2012}%
  \BibitemOpen
  \bibfield  {author} {\bibinfo {author} {\bibfnamefont {H.}~\bibnamefont
  {Shinaoka}}, \bibinfo {author} {\bibfnamefont {T.}~\bibnamefont {Miyake}}, \
  and\ \bibinfo {author} {\bibfnamefont {S.}~\bibnamefont {Ishibashi}},\ }\href
  {\doibase 10.1103/PhysRevLett.108.247204} {\bibfield  {journal} {\bibinfo
  {journal} {Physical Review Letters}\ }\textbf {\bibinfo {volume} {108}},\
  \bibinfo {pages} {247204} (\bibinfo {year} {2012})}\BibitemShut {NoStop}%
\bibitem [{\citenamefont {Aryasetiawan}\ \emph {et~al.}(2004)\citenamefont
  {Aryasetiawan}, \citenamefont {Imada}, \citenamefont {Georges}, \citenamefont
  {Kotliar}, \citenamefont {Biermann},\ and\ \citenamefont
  {Lichtenstein}}]{aryasetiawan_frequency-dependent_2004}%
  \BibitemOpen
  \bibfield  {author} {\bibinfo {author} {\bibfnamefont {F.}~\bibnamefont
  {Aryasetiawan}}, \bibinfo {author} {\bibfnamefont {M.}~\bibnamefont {Imada}},
  \bibinfo {author} {\bibfnamefont {A.}~\bibnamefont {Georges}}, \bibinfo
  {author} {\bibfnamefont {G.}~\bibnamefont {Kotliar}}, \bibinfo {author}
  {\bibfnamefont {S.}~\bibnamefont {Biermann}}, \ and\ \bibinfo {author}
  {\bibfnamefont {A.~I.}\ \bibnamefont {Lichtenstein}},\ }\href {\doibase
  10.1103/PhysRevB.70.195104} {\bibfield  {journal} {\bibinfo  {journal}
  {Physical Review B}\ }\textbf {\bibinfo {volume} {70}},\ \bibinfo {pages}
  {195104} (\bibinfo {year} {2004})}\BibitemShut {NoStop}%
\bibitem [{\citenamefont {Liu}\ \emph {et~al.}(2020)\citenamefont {Liu},
  \citenamefont {He}, \citenamefont {Kim}, \citenamefont {Khmelevskyi},
  \citenamefont {Toschi}, \citenamefont {Kresse},\ and\ \citenamefont
  {Franchini}}]{liu_comparative_2020}%
  \BibitemOpen
  \bibfield  {author} {\bibinfo {author} {\bibfnamefont {P.}~\bibnamefont
  {Liu}}, \bibinfo {author} {\bibfnamefont {J.}~\bibnamefont {He}}, \bibinfo
  {author} {\bibfnamefont {B.}~\bibnamefont {Kim}}, \bibinfo {author}
  {\bibfnamefont {S.}~\bibnamefont {Khmelevskyi}}, \bibinfo {author}
  {\bibfnamefont {A.}~\bibnamefont {Toschi}}, \bibinfo {author} {\bibfnamefont
  {G.}~\bibnamefont {Kresse}}, \ and\ \bibinfo {author} {\bibfnamefont
  {C.}~\bibnamefont {Franchini}},\ }\href {\doibase
  10.1103/PhysRevMaterials.4.045001} {\bibfield  {journal} {\bibinfo  {journal}
  {Physical Review Materials}\ }\textbf {\bibinfo {volume} {4}},\ \bibinfo
  {pages} {045001} (\bibinfo {year} {2020})}\BibitemShut {NoStop}%
\end{thebibliography}%
\end{document}